\documentclass[aps,12pt,eqsecnum, preprint,nofootinbib,superscriptaddress]{revtex4}
\usepackage{amssymb,amsmath,amsthm,graphicx,amscd,mathtools}
\usepackage{upgreek,enumerate,color,verbatim,comment,ulem,stackrel}
\usepackage{bm}

\usepackage[hidelinks]{hyperref}

\begin{document}

\title{Quantum Thermodynamic Uncertainties\\  in Nonequilibrium Systems\\ from Robertson-Schr\"odinger Relations}
\author{Hang Dong}
\affiliation{Center for Field Theory and Particle Physics, Department of Physics, Fudan University, Shanghai 200433, People's Republic of China}
\author{Daniel Reiche}
\affiliation{Humboldt-Universit\"at zu Berlin, Newtonstr. 15, 12489 Berlin, Germany}
\author{Jen-Tsung Hsiang}
\email{cosmology@gmail.com}
\affiliation{Center for High Energy and High Field Physics, National Central University, Taoyuan 320317, Republic of China}
\author{Bei-Lok Hu}
\email{blhu@umd.edu}
\affiliation{Maryland Center for Fundamental Physics and Joint Quantum Institute,  University of Maryland, College Park, MD 20742, USA}


\date{\today}

\begin{abstract}
Thermodynamic uncertainty principles make up one of the few rare anchors in the largely uncharted waters of nonequilibrium systems, the fluctuation theorems being the more familiar.
In this work we aim to trace the uncertainties of thermodynamic quantities in nonequilibrium systems to their quantum origins, namely, to the quantum uncertainty principles.  Our results enable us to make this categorical statement: For Gaussian systems, thermodynamic functions are functionals of the {\it Robertson-Schr\"odinger uncertainty} function, which is always non-negative for quantum systems, but not necessarily so for  classical systems.  Here, quantum refers to noncommutativity of the canonical operator pairs. From the {\it nonequilibrium free energy}~\cite{hsiang21}, we succeeded in deriving several inequalities between certain thermodynamic quantities. They assume the same forms as those in conventional thermodynamics, but these are nonequilibrium in nature and they hold for all times and at strong coupling. In addition we show that a fluctuation-dissipation inequality exists at all times in the nonequilibrium dynamics  of the system. For nonequilibrium systems which relax to an equilibrium state at late times, this fluctuation-dissipation inequality leads to the Robertson-Schr\"odinger uncertainty principle with the help of the Cauchy-Schwarz inequality. This work provides the microscopic quantum basis to certain important thermodynamic properties of macroscopic nonequilibrium systems.
\end{abstract}
\maketitle

\tableofcontents
\clearpage
\baselineskip=18pt

\numberwithin{equation}{section}
\allowdisplaybreaks

\section{Introduction}

Uncertainty in {simultaneous} measurements of canonical variables is one of the building blocks of modern quantum field {theories.}
Already in the early days of quantum mechanics, they found their formalization in, e.g., the Heisenberg Uncertainty principle \cite{heisenberg27} for zero temperature systems, the Robertson-Schr\"odinger uncertainty function \cite{robertson29,schroedinger30} for fully nonequilibrium quantum systems or in more recently developed relations for Markovian \cite{lindblad76,lu89,sandulescu87} and non-Markovian \cite{hu93a,anastopoulos95,hu95,koks97} open quantum systems (see also, e.g., Ref. \cite{kempf95} for a generalization with non-zero minimal uncertainty in one variable).
The uncertainty principle at finite temperature has been shown (e.g., \cite{hu93a,hu95}) to be a  useful indicator and a valid alternative to decoherence criteria \cite{zurek03,joos03,schlosshauer08a}) of quantum to classical transition, in that a crossover temperature can be identified between a vacuum fluctuations-dominated regime at very low temperatures to a
thermal fluctuations-dominated high temperature regime where conventional thermodynamics
applies.
Such relations are often motivated by a microscopic perspective on the dynamical properties of the single quantum degrees of freedom.

While it might be no surprise that uncertainty relations will leave their marks on macroscopic (thermodynamic) parameters of the system, general statements in nonequilibrium systems are {scarce}.
Importantly, we should mention entropic uncertainty relations \cite{coles17}, fluctuation theorems \cite{jarzynski96,crooks99}, and the recently developed thermodynamic uncertainty relations {\cite{horowitz20}}.
Also, an extensive effort has been made  to find generalizations (see, e.g., the fluctuation-dissipation inequality \cite{fleming13}) as well as extensions (see, e.g., local thermal equilibrium \cite{polder71,eckhardt84}) of thermodynamic equilibrium results to nonequilibrium {systems}.
The exact {lineage}, however, from (microscopic) uncertainty relations to (macroscopic) quantum thermodynamic uncertainties in nonequilibrium situations is, to the best of our knowledge, not completely transparent yet.

\subsection{Our Intents, modeling and methodology}
In our work, we intend to approach this issue in two stages.
In this first paper, we do not adhere to any specific formulation or particular interpretation  of thermodynamic uncertainties such as offered in the above references, but explore and strengthen their theoretical foundations using the conceptual framework of open quantum systems (OQS) \cite{Weiss,breuer07,RivHue} and the tools of nonequilibrium (NEq) quantum field theory  \cite{rammer07,calzetta08,kamenev11}.
Our goal is to make explicit the mechanical and dynamical basis of all thermodynamic quantities or relations. In concrete terms, for the purpose stated here, especially for quantum systems, this means tracing the uncertainties of thermodynamic quantities all the way to their origins, to the (microscopic) quantum uncertainty principles (QUP).

In future work \cite{QTUR2}, we shall formulate specific uncertainty relations for thermodynamic quantities based on the results of this paper.
Using the methodology of NEq quantum dynamics, we thereby provide some notable formulations in the literature with the connection to their microscopic foundations.
Our results will be of relevance for a number of active areas of current research, such as exploring the existence of fluctuation-dissipation relations (or inequalities) and thermodynamic uncertainty relations in quantum friction \cite{reiche22}, the dynamical Casimir effect \cite{dodonov20} or quantum processes in the early universe \cite{hu20}.

Before describing our modeling and methodology, we wish to first mention the linkage with some earlier works by two of us on this topic, which will serve as the theoretical basis for the present work.
In conventional thermodynamics, the fluctuation-dissipation relations (FDR) \cite{callen51,kubo66,li93} are usually derived and explained in the context of linear response theory (LRT), where the behavior of linear perturbations to a system in equilibrium with a bath is studied, the former manifesting as dissipative dynamics and the latter as noise.
In the context of fully nonequilibrium dynamics, except for very special situations where mathematical theorems can be proven, one needs to know explicitly and follow the evolution of the system while it interacts with its environment to late times,  and examine if conditions exist for the system to fully equilibrate.
Only {\it after} this equilibrium condition is met can one  explore whether such relations exist or not.
In the recent works by two of us \cite{hsiang15,hsiang18,hsiang19,hsiang20a,hsiang20b}, we explicitly calculated the power balance between the system and its thermal bath(s) to show that indeed the FDRs are in place.
For other approaches to this theme, see, e.g.,  \cite{polevoi75,eckhardt84,pottier01,seifert10,seifert12,barton16,lopez18,sinha21}.
Moving away from the near-equilibrium conditions of LRT where FDRs are expected, to the fully nonequilibrium conditions,  where there are no FDRs a priori, it is natural to speculate whether certain time-dependent {fluctuation-dissipation} inequalities may exist, which tag along the system's nonequilibrium evolution, changing over to bona fide FDRs at late times,  and,  whether some bounds can be deduced which cover the whole course of the system's temporal development.
In lieu of exact relations similar to the equilibrium FDR, the present work will rely on two principles of linear open quantum systems.
First, imposing a few physical demands, i.e. that the interaction is causal and the respective operators describing system and environment are hermitian, a fluctuation-dissipation \textit{inequality} (FDI) can be {constructed} to hold for the full course of the interaction \cite{fleming13} (see Sec.~\ref{Sec:FDIandRS} for the connection between the FDI and the FDR and Ref. \cite{reiche20a} for a useful application to the nonequilibrium thermodynamics of quantum friction).
Second, a quantum uncertainty function can be identified for Gaussian open quantum systems {that is related to} the recently {described} nonequilibrium free energy density and the associated partition functions \cite{hsiang21} (see Secs. \ref{Sec:NEqFreeEnergy} and \ref{Sec:RSandUncertaintyFunction}).

The model we use is the generic quantum Brownian motion (QBM) model \cite{schwinger61,feynman63,caldeira83,grabert88,hu92,hu93b}, with a harmonic oscillator as the system and a finite temperature scalar field as its environment,  covering the full temperature range.  Thus we have in mind using the Hu-Paz-Zhang (HPZ) master equation \cite{hu92,hu93b} or its equivalent Fokker-Planck-Wigner \cite{halliwell96} or the Langevin \cite{ford88,calzetta01,calzetta03} equations as templates, for the description of its non-Markovian (back-action incorporated, self-consistent) dynamics. Alternatively we shall also use the covariant matrix elements~\cite{ai07} {derived from a set of Langevin equations} to describe the nonequilibrium  dynamics of the reduced quantum system.
The goal is to obtain uncertainty relations for quantum thermodynamic quantities and FDIs for fully nonequilibrium dynamics.
We note that, since many popular approximations (such as the Born-Markov or the rotating wave approximation) give incomplete or even wrong results in the full regimes we want to explore, and we do not want the true nature of thermodynamic relations or bounds in inequalities to be affected by them, we set forth to find exact solutions. This is why we prefer to work with Gaussian open quantum systems in addressing basic theoretical issues.

\subsection{Our findings in relation to background works }
\begin{enumerate}
	\item This work aims at exploring the relationship between thermodynamic and quantum uncertainty principles, as well as the existence and meaning of fluctuation-dissipation inequalities. Related to this work are two groups of our earlier papers :
	\begin{enumerate}[a)]
		\item The {\it uncertainty principle at finite temperature} has been shown (e.g.,~\cite{hu93a,hu95}) to be a useful indicator of quantum to classical transition, in that a crossover temperature can be identified between a vacuum fluctuations-dominated regime at very low temperatures to a thermal fluctuations-dominated high temperature regime where conventional thermodynamics applies.
		\item A {\it quantum fluctuation-dissipation inequality} exists in a thermal quantum bath~\cite{fleming13}: quantum fluctuations are bounded below by quantum dissipation, whereas classically the fluctuations vanish at zero temperature. The lower bound of this inequality is exactly satisfied by (zero-temperature) quantum noise and is in accord with the Heisenberg uncertainty principle. This inequality has been applied to understand issues in quantum friction~\cite{reiche20a} (see, e.g.,~\cite{reiche22} for background).
A good summary of recent work on the relation of thermodynamic uncertainty relations and non-equilibrium fluctuations can be found in, e.g.,~\cite{horowitz20} and references therein.
	\end{enumerate}
	\item Toward our stated goals, as a preamble, we can make this categorical statement: For Gaussian systems, thermodynamic functions are functionals of the {\it Robertson-Schr\"odinger uncertainty} (RSU) function \eqref{E:kghdbfg}, which is always non-negative for quantum systems, but not necessarily so for  classical systems.  Here, quantum refers to noncommutativity of the canonical operator pairs.
	\item The expectation value of the {\it nonequilibrium Hamiltonian of mean force}~\cite{hsiang21} gives the nonequilibrium internal energy of the system, and is bounded from above by the expectation value of the system's Hamiltonian.
	 \item The {\it nonequilibrium heat capacity}, derived from taking the derivative of the nonequilibrium internal energy with respect to the {\it nonequilibrium effective temperature},  remains proportional to the fluctuations of nonequilibrium Hamiltonian of mean force,  with a proportionality constant   given by the nonequilibrium effective temperature, not the bath temperature. These results apply for all times and at strong coupling.
	\item From the {\it nonequilibrium free energy}~\cite{hsiang21}, we succeeded in deriving several inequalities (\eqref{E:uutgkh} and \eqref{E:lejori}) between certain thermodynamic quantities. They assume the same forms as those in conventional thermodynamics, but emphatically,  these are nonequilibrium in nature and  they hold for all times and at strong coupling.
	\item Fluctuation-dissipation inequalities (FDI) and relation with fluctuation-dissipation relation (FDR)
	\begin{enumerate}[a)]
		\item For those systems that can reach stationary states at late times, it was shown earlier ~\cite{callen51,kubo66,li93,hsiang15,hsiang18,hsiang19,hsiang20a,hsiang20b} that they possess a fluctuation-dissipation relation.
	\item We have shown that a fluctuation-dissipation inequality exists at all times in the nonequilibrium dynamics of the system.
	\item At late times in the nonequilibrium relaxation of the reduced system, this fluctuation-dissipation inequality leads to the Robertson-Schr\"odinger uncertainty principle with the help of the Cauchy-Schwarz inequality.
	\end{enumerate}
	\item While the mathematical expressions of FDI have been found (6-b), we want to further understand their physical meanings. While the relation between FDI and RSU has been found (6-c) in the stage when the reduced system are closely approaching equilibrium, we want to find out whether there is a connection between the FDIs and the FDRs.
	This will provide a useful linkage between the more challenging nonequilibrium dynamics and the more familiar equilibrium states.
\end{enumerate}

{The paper is structured as follows.
In Sec. \ref{Sec:NEqFreeEnergy}, we derive the dynamical behavior of the covariance matrix and review the formalism of the nonequilibirum free energy.
We further highlight the special role of the Robertson-Schrödinger uncertainty function for Gaussian systems.
In Sec. \ref{Sec:NEqThermodynamic}, we establish a connection between the system's internal energy and the expectation value of the system's Hamiltonian.
{We} derive the fluctuation-dissipation inequality and explore its connection to the Robertson-Schr\"odinger uncertatinty function as well as its dynamical behavior when approaching equilibrium {in Sec. \ref{Sec:FDIandRS}}.
We conclude our work with a discussion in Sec. \ref{Sec:Conclusion}.}

\section{Nonequilibrium dynamics of Gaussian open quantum systems}\label{Sec:NEqFreeEnergy}

{We begin our discussion}, for the convenience of the reader, {by summarizing} the key features of the nonequilibrium dynamics of Gaussian {open} quantum systems {and recapitulate the method of the nonequilibrium free energy introduced in Ref.} \cite{hsiang21}.
{Throughout the manuscript, we use units such that $\hbar=k_{\textsc{b}}=1$.}

The specific system under consider is a point-like object, with internal degrees of freedom modeled by a quantum harmonic oscillator.
This could, for example, correspond to the physical systems of an Unruh-deWitt detector~\cite{hsiang21a} or the (electric) dipole resonance of a neutral atom~\cite{intravaia11a}.
The external (spatial) degrees of freedom of the system, for simplicity, are assumed to be fixed at the origin of space.
We use a massless quantum scalar field that is initially in its thermal state to act as a thermal bath.
We further assume the coupling between the system and the bath to be linear with respect to the oscillator's displacement and the field variables.
The coupling strength between system and bath, however, is left arbitrarily strong.

If the initial state of the system is {Gaussian}, the reduced state of the system is guaranteed to remain Gaussian.
At any moment, the Gaussian state in general can be decomposed as
\begin{equation}\label{E:gfnksbs}
	\hat{\rho}=\hat{D}(\alpha)\hat{S}(\zeta)\hat{R}(\theta)\,\hat{\rho}_{\vartheta}\,\hat{R}^{\dagger}(\theta)\hat{S}^{\dagger}(\zeta)\hat{D}^{\dagger}(\alpha)\,,
\end{equation}
{where we defined} the squeeze operator $\hat{S}(\zeta)$, the displacement operator $\hat{D}(\alpha)$, and the rotation operator $\hat{R}(\theta)$.
Here, $\hat{\rho}_{\vartheta}$ is a Gibbs state of the form
\begin{align}\label{E:rjugfg}
	\hat{\rho}_{\vartheta}&=\frac{1}{\mathcal{Z}_{\vartheta}}\,\exp\Bigl[-\vartheta\bigl(\hat{a}^{\dagger}\hat{a}+\frac{1}{2}\bigr)\Bigr]\,, &\mathcal{Z}_{\vartheta}&=\operatorname{Tr}\exp\Bigl[-\vartheta\bigl(\hat{a}^{\dagger}\hat{a}+\frac{1}{2}\bigr)\Bigr]\,, &\vartheta&\geq0\,,
\end{align}
in which the Gibbs parameter $\vartheta=\vartheta(T_{\textsc{b}})$ depends on the initial bath temperature {$T_{\textsc{b}}$}, and $\hat{a}$ {($\hat{a}^{\dagger}$) is} the creation {(annihilation)} operator associated with the quantum harmonic oscillator, satisfying the standard commutation relation $[\hat{a},\hat{a}^{\dagger}]=1$.
The squeeze operator $\hat{S}(\zeta)$ takes the form
\begin{align}
	\hat{S}(\zeta)&=\exp\biggl[\frac{1}{2}\,\zeta^{*}a^{2}-\frac{1}{2}\,\zeta\,a^{\dagger\,2}\biggr]\,,&\zeta&\in\mathbb{C}\,,
\end{align}
{with the} squeeze parameter {$\zeta=\eta\,e^{i\phi}$ ($0\leq\eta<\infty$ and $0\leq\theta<2\pi$)}.
For simplicity, we are interested in a situation where the harmonic quantum system is initially in its ground state such that the displacement operator vanishes for our configuration. Further, as the rotation operator describes a global phase only, it can be ignored by simply setting $\theta=0$ without loss of generality.
Squeezing here is the consequence of finite coupling between system and bath~ \cite{intravaia03,maniscalco04,hsiang21}.

Due to the choice for the system's initial state, the first moments of the canonical variable operators vanish.
The initial state of the combined system is in general not an eigenstate of the total Hamiltonian, which comprises the system, the bath and the interaction Hamiltonian.
{While the} state of the combined systems will unitarily evolve with time, the reduced state of the system {[Eq. \eqref{E:gfnksbs}]}, {in turn}, will evolve  non-unitarily with time. This implies {that} the parameters in \eqref{E:gfnksbs}, in particular the squeeze parameter $\zeta$ and the Gibbs parameter $\vartheta$, are functions of time.
{Even though this can lead to a quite involved form of the density matrix at arbitrary times, a} convenient feature of the decomposition {in} \eqref{E:gfnksbs} {is the invariance of the trace with respect to squeezing, i.e.}
\begin{equation}\label{E:kgkdtr}
	\operatorname{Tr}\Bigl\{\hat{S}(\zeta)\,\exp\Bigl[-\vartheta\bigl(\hat{a}^{\dagger}\hat{a}+\frac{1}{2}\bigr)\Bigr]\,\hat{S}^{\dagger}(\zeta)\Bigr\}=\operatorname{Tr}\exp\Bigl[-\vartheta\bigl(\hat{a}^{\dagger}\hat{a}+\frac{1}{2}\bigr)\Bigr]=\mathcal{Z}_{\vartheta}\,.
\end{equation}
This suggests that $\mathcal{Z}_{\vartheta}$ can serve as a ({nonequilibrium}) partition function associated with the state \eqref{E:gfnksbs}.
It is uniquely fixed by the parameter $\vartheta$.
A keen reader may wonder why the partition function does not depend on the squeeze parameter $\zeta$, which carries the dynamics of the reduced system. We will justify it later in Sec.~\ref{Sec:NEqPart} when we discuss the role of the Robertson-Schr\"odinger inequality in quantum thermodynamics.
Hereafter we will assign $\mathcal{Z}_{\vartheta}$ as the nonequilibrium partition function $\mathcal{Z}_{\textsc{s}}$ of the reduced system, from which the nonequilibrium free energy, and in turn, various thermodynamic functions can be introduced.
{We} will now connect the nonequilibrium dynamics with the nonequilibrium thermodynamics of the reduced system.

\subsection{{Dynamical behavior of the} covariance matrix}\label{Sec:DynB}

{Due to our choice of} the initial states of the reduced system {(vanishing first moments of the canonical variable's operators)}, we {have seen in the previous section that only a set of  three parameters is needed} to characterize the {complete}  dynamics {of} the reduced system.
{Although $\{\vartheta,\eta,\phi\}$ is a convenient choice to construct the nonequilibrium partition function, another, perhaps more frequently used, set} of parameters {is encoded in} the elements of the covariance matrix $\bm{C}(t)=\frac{1}{2}\langle\{\hat{\bm{R}},\hat{\bm{R}}^{T}\}\rangle$, with $\hat{\bm{R}}^{T}=(\hat{Q},\hat{P})$ in which $\hat{P}$ is the momentum conjugated  to the displacement $\hat{Q}$
of the quantum harmonic oscillator\footnote{{Note that the} covariance matrix $\bm{C}$ {more generally} is defined in terms of {its variance} $\Delta\hat{\bm{R}}\equiv\hat{\bm{R}}-\langle\hat{\bm{R}}\rangle$, instead of $\hat{\bm{R}}$. However, {for our purposes with vanishing first moments, this} makes no difference.}.
The expectation value $\langle\cdots\rangle$ is defined with respect to the density matrix operator \eqref{E:gfnksbs}, and $\{\cdots\}$ is the anti-commutator.
The covariance matrix contains three linearly independent functions of time which serve as alternative set of parameters $\{a,b,c\}$, i.e.
\begin{align}
	a(t)&=\langle\hat{P}^{2}(t)\rangle\,,&b(t)&=\langle\hat{Q}^{2}(t)\rangle\,,&c(t)&=\frac{1}{2}\langle\bigl\{\hat{Q}(t),\,\hat{P}(t)\bigr\}\rangle\,.
\end{align}
{For the exact mapping between the sets $\{\vartheta,\eta,\phi\}\to\{a,b,c\}$, we refer, e.g., to Ref. \cite{hsiang21} (see also Sec. \ref{Sec:NEqThermodynamic}).}
Both sets are equivalent and can be interchanged for convenience.
The determinant of the covariance matrix elements can be used to express the Robertson-Schr\"odinger uncertainty principle,
\begin{align}
	ab-c^{2}
	=
	\langle
	\hat{Q}^{2}(t)\rangle\langle\hat{P}^{2}(t)
	\rangle
	-\frac{1}{4}
	\langle
	\bigl\{\hat{Q}(t),\,\hat{P}(t)\bigr\}
	\rangle^{2}
	&\geq
	\frac{1}{4}
	\,.\label{E:fjghdsg}
\end{align}
In principle we may always re-define the canonical variables to make $c$ vanish.
{However,} in the context of nonequilibrium evolution of the reduced open system, this re-definition is time-dependent, so we will {simply} fix the choice of canonical variables.
{In this way, }$c=0$ has {the} special physical significance {of indicating the equilibration between system and environment (see, e.g., Refs. \cite{ford01,fleming11} and also below)}.

The dynamics of the covariance matrix elements have been extensively discussed in the literature.
Here we will cite a few essential properties.
Recall that the field is evaluated at spatial origin, where the oscillator is located, so we suppress the corresponding spatial dependencies in the quantities.
The covariance elements are then given by the response to the free field's (Hadamard) Green function $G_{\textsc{h}}^{(\phi)}$, i.e.
\begin{subequations}
\begin{align}
a(t)&=m^{2}\dot{d}_{1}^{2}(t)\,\langle\hat{Q}^{2}(0)\rangle+\dot{d}_{2}^{2}(t)\,\langle\hat{P}^{2}(0)\rangle+e^{2}\int_{0}^{t}\!{\mathrm{d}s\mathrm{d}s'}\;\dot{d}_{2}(t-s)\dot{d}_{2}(t-s')\,G_{\textsc{h}}^{(\phi)}{(s,s')}\,,\label{E:fnghj}\\
	b(t)&=d_{1}^{2}(t)\,\langle\hat{Q}^{2}(0)\rangle+\frac{d_{2}^{2}(t)}{m^{2}}\,\langle\hat{P}^{2}(0)\rangle+\frac{e^{2}}{m^{2}}\int_{0}^{t}\!{\mathrm{d}s\mathrm{d}s'}\;d_{2}(t-s)d_{2}(t-s')\,G_{\textsc{{h}}}^{(\phi)}{(s,s')}\,,\label{E:gfjbskgf1}\\
	c(t)&=m\,d_{1}(t)\dot{d}_{1}(t)\,\langle\hat{Q}^{2}(0)\rangle+\frac{d_{2}(t)\dot{d}_{2}(t)}{m}\,\langle\hat{P}^{2}(0)\rangle+\frac{e^{2}}{m}\int_{0}^{t}\!{\mathrm{d}s\mathrm{d}s'}\;d_{2}(t-s)\dot{d}_{2}(t-s')\,G_{\textsc{h}}^{(\phi)}{(s,s')}\,,\label{E:gfjbskgf3}
\end{align}
\end{subequations}
where $d_{1}(t)$, $d_{2}(t)$, satisfying
\begin{align}
	d_{1}(0)&=1\,,&\dot{d}_{1}(0)&=0\,,&d_{2}(0)&=0\,,&\dot{d}_{2}(0)&=1\,,
\end{align}
are a special set of homogeneous solutions to the quantum Langevin equation
\begin{equation}
	\ddot{\hat{Q}}(t)+2\gamma\,\dot{\hat{Q}}(t)+\omega_{\textsc{r}}^{2}\hat{Q}(t)=\frac{e}{m}\,\hat{\phi}{(t)}\,.\label{E:gkfha}
\end{equation}
{Also,} $m$ is the mass of the quantum harmonic oscillator, $e$ is the coupling strength between the system (oscillator) and the bath (free scalar field $\hat{\phi}$), $\omega_{\textsc{r}}$ is the physical frequency of the oscillator, and $\gamma=e^{2}/(8\pi m)$ is the damping constant.
The nonlocal version of the Langevin equation \eqref{E:gkfha} can  be obtained by solving a simultaneous set of Heisenberg equations {for} the oscillator and the field{, respectively}.
It reduces to the local form \eqref{E:gkfha} for the massless quantum field we choose.
{We note that,} if the state of the system at initial time $t=0$ has a nonvanishing correlation between the pair of canonical operators, then the expressions \eqref{E:fnghj}--\eqref{E:gfjbskgf3} for the covariance matrix elements will have an additional term depending on this initial correlation.

The effects of the environment {(bath)} are embedded in the Hadamard function $G_{\textsc{h}}^{(\phi)}$,  which describes the {fluctuations} of the scalar field.
It is defined as (at $\bm{x}=\bm{x}'=0$)
\begin{equation}
	G_{\textsc{h}}^{(\phi)}(t,t')=\frac{1}{2}\operatorname{Tr}\biggl(\hat{\rho}_{\textsc{b}}\bigl\{\hat{\phi}(t),\,\hat{\phi}(t')\bigr\}\biggr)\,
\end{equation}
{and hence} depends on the initial state $\hat{\rho}_{\textsc{b}}$ of the environment.
It quantifies the influence of quantum field fluctuations on the reduced system.
{This can be seen from the Langevin equation in Eq.~\eqref{E:gkfha}:} $\hat{\phi}$ serves as a quantum noise force, driving the dynamics of the reduced system via $\hat{Q}$. Accompanying this noise force, owing to the system-bath interaction,  dissipative effects  show  up in the reduced system dynamics.
It is concealed in $d_{1}(t)$ and $d_{2}(t)$  which decay exponentially with time.

At equilibrium, the fluctuations and the dissipation effects of the environment are related by a fluctuation-dissipation relation.
This relation gauges precisely the energy exchange between the system and the environment if the dynamics of the reduced system can come to equilibration at late times. In the special case of vanishingly weak system-bath interaction, the equilibration becomes thermalization in the context of open systems.
Otherwise, in general, the final equilibrium state of the reduced system does not have a Gibbs form.
{From this perspective, the possibility of equilibration of the reduced system as a result of its nonequilibrium evolution can be paraphrased as the existence of the fluctuation-dissipation relation of the system.}
Before the system comes to equilibrium, although there is no fluctuation-dissipation relation, we can nevertheless derive a fluctuation-dissipation inequality (FDI).
This, and the role the FDI plays in the thermodynamic {inequalities/uncertainties, will be the main theme of the remainder of this} and our subsequent paper.

As the structures of the covariance matrix elements is important, we make a few observations on their generic behavior:
\begin{enumerate}
	\item Each element can be divided into an active and a passive component.
	The active component depends on the initial state of the system and represents the intrinsic quantum nature of the system.
	The passive component relies on the initial state of the environment and represents the induced quantum effects of the environment~\cite{calzetta08}.
	\item Since $d_{1}(t)$, $d_{2}(t)$ decay with time, the active component will diminish at late times. The behavior of the covariance matrix elements at late times are essentially governed by the environment.
	{In this way,} the statistical and causality properties of the environment will eventually be passed on to the reduced system.
	\item The damping term in Eq.~\eqref{E:gkfha} is proportional to the ``velocity'' of $\hat{Q}$, {i.e. the canonical momentum.}
	{Hence,} if the reduced system has a small initial momentum uncertainty, then we expect that damping plays a minor role compared to the noise force {at the early stage of the nonequilibrium evolution}.
	{Accordingly,} the elements $a$, $b$ will increase with time, mainly driven by the fluctuation force. In due course, the momentum uncertainty will grow sufficiently large {such} that the damping effect gradually catches up with the noise effect.
	\item In contrast, if the reduced system has a large initial momentum uncertainty, the damping effect will start  off strongly, and the noise effect is subdominant. The elements $a$, $b$  decrease with time until the damping effect is small enough to match up with the effect of the noise force.
	\item The element $c$ does not have a definite sign, but oscillates with time. However,  from parity consideration, when equilibrium is reached, the time- translational  invariance of the state requires that $c$ should vanish. Thus, {$c=0$} may serve as an indicator of the existence of an asymptotic equilibrium state. In the final equilibrium state, both $a$ and $b$ are constant in time.
	\item The Hadamard function of the bath $G_{\textsc{h}}^{(\phi)}$ can be decomposed into two contributions in the current setting: One results from vacuum fluctuations of the field,  the other from thermal fluctuations.
	{We note that additional terms would appear if macroscopic bodies were present in the surroundings of the particle, which would carry their own (material-modified) quantum and thermal fluctuations \cite{rytov53}.}
	At low bath temperatures $\beta_{\textsc{b}}\omega_{\textsc{r}}\gtrsim1$, the vacuum fluctuation effects dominate, while at higher bath temperatures, they become  insignificant. The vacuum fluctuations of the massless field are scaleless, {such that the Hadamard function} has a rather long range effect {at the order of the squared inverse distance from the source}.
	{Again, if additionally material bodies were present, the situation can be different in the vicinity of surfaces where, depending on the structure of the material \cite{carminati15,woods16}, often higher-order inverse polynonmials \cite{intravaia11a,oelschlaeger18,reiche19} or even non-algebraic functions \cite{reiche17,reiche20} for the distance-dependence occur.}
	\item The previous discussions and formalism are not restricted to weak coupling.
	They also apply to the case of strong coupling $\gamma/\omega_{\textsc{r}}\sim\mathcal{O}(1)$, as long as the dynamics is stable.
	For strong coupling, since the scales of the reduced system like $\gamma$, $\omega_{\textsc{r}}$, and the bath temperature $\beta_{\textsc{b}}^{-1}$ can {become} comparable in magnitude, the curves for the temporal evolution of {the covariance elements} of the reduced system {can show a rich structure during the nonequilibrium evolution (see, e.g., Fig. \ref{Fi:UnCent})}.
\end{enumerate}

\subsection{Robertson-Schr\"odinger uncertainty principle}\label{Sec:RSandUncertaintyFunction}

Since the Robertson-Schr\"odinger uncertainty principle can be expressed in terms of  the covariance matrix elements, as shown in \eqref{E:fjghdsg}, we expect that {its generic} behavior {during the nonequilibrium dynamics} will be passed on to the uncertainty principle, revealing finer details about the uncertainty principle rather than a monotonous inequality.
To this end, {it is convenient} to introduce the uncertainty function
\begin{equation}\label{E:kghdbfg}
	\mathfrak{S}(t)=a(t)b(t)-c^{2}(t)-\frac{1}{4}\,,
\end{equation}
which is {positive semi-}definite if the Robertson-Schr\"odinger uncertainty principle is satisfied.
However, for quantum open systems with the quantum field as one of the subsystems, it {can be} nontrivial to realize the positive semi-definiteness due to the presence of the cutoff and the implementation of the renormalization schemes.
%
{
For instance, the term $a=\langle\hat{P}^{2}\rangle$, when compared to the autocorrelation of the position operator, carries extra time derivatives and the corresponding extra factor of $\omega^2$ in the high-frequency domain requires a proper regularization to tame the UV-divergence.
Hence, simply dropping the cutoff-dependent expressions in the covariance matrix elements, in particular {$a=\langle\hat{P}^{2}\rangle$}, may result in serious inconsistencies such as the uncertainty function becoming negative.}
{For the present case,} in the numerical calculations, we have inserted a convergence factor of the form $e^{-\kappa\epsilon}$ in the integrands whenever it is necessary to make the integrals over the frequency $\kappa$ well defined.
We typically choose $\epsilon$ to be $\omega_{\textsc{r}}/20$.
We will postpone the discussion about the consequence of the cutoff scale to Sec.~\ref{Sec:NEqPart} in the context of the effective temperature.

\begin{figure}
\centering
    \scalebox{0.4}{\includegraphics{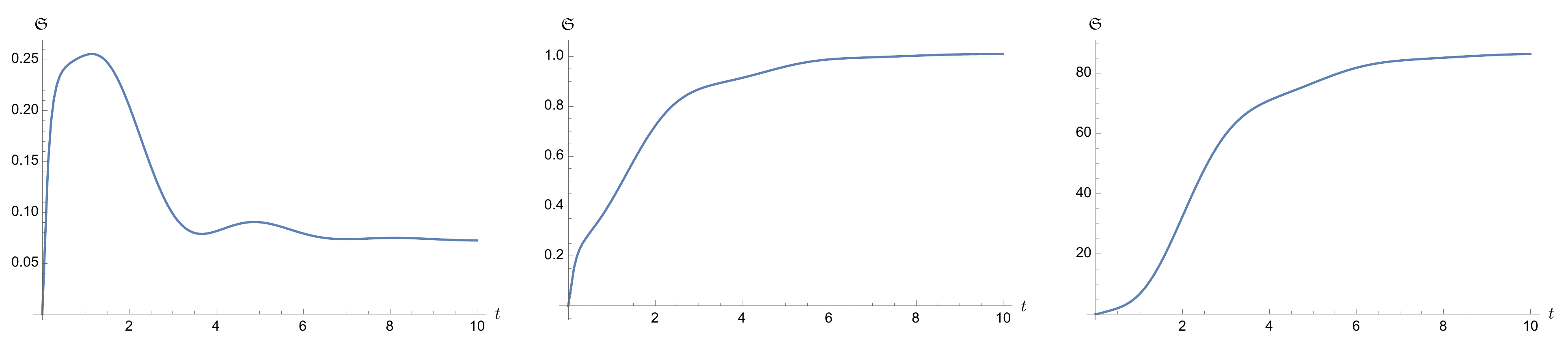}}
    \caption{The time dependence of the uncertainty function $\mathfrak{S}(t)$ is plotted for three different bath temperatures. From the left to the right, the inverse bath temperature $\beta_{\textsc{b}}$ is 10, 1, 0.1, normalized with respect to $\omega_{\textsc{r}}^{-1}$.
		The damping constant $\gamma$ is $0.3\times\omega_{\textsc{r}}$ and the oscillator mass $m$ is $m=1\times\omega_{\textsc{r}}$. In the low temperature regime $\beta_{\textsc{b}}\omega_{\textsc{r}}>1$, the finite temperature contribution is subdominant, and the effects due to vacuum fluctuations of the bath and its cutoff are more prominent.
		}
		\label{Fi:UnCent}
\end{figure}

We report on a numerical evaluation of the uncertainty function in Fig.~\ref{Fi:UnCent}.
As expected, the uncertainty function is always non-negative.
Since we choose the initial state of the system to be the ground state of the oscillator, the uncertainty function always arises from zero but gradually saturates to a constant at late times, due to the equilibration of the reduced dynamics.
At lower temperatures $\beta_{\textsc{b}}^{-1}<\omega_{\textsc{r}}$ (left plot of Fig.~\ref{Fi:UnCent}), a ``bump'' appears at times before relaxation.
This results from two factors: {First,} in this regime, the thermal fluctuations of the bath give a minor contribution, compared to the corresponding vacuum fluctuations.
Thus the cutoff-dependent contribution from the quantum field bath stands out.
Second, it in turn contributes to rapid rising of momentum, and hence the uncertainty function overshoots {its late-time value}. It then quickly decays to a constant due to the relaxation process caused by the damping in the reduced dynamics.
At sufficiently high bath temperatures (right plot of Fig. \ref{Fi:UnCent}), the thermal fluctuations dominate the interaction, as seen by the absolute magnitude of the uncertainty function at late times.
The complicated behavior seen in the low-temperature situation (left plot of Fig. \ref{Fi:UnCent}) is overshadowed by the thermal dynamics.
For comparison, we show the intermediate regime where $\beta_{\textsc{b}}^{-1}=\omega_{\textsc{r}}$ (center plot of Fig. \ref{Fi:UnCent}).
The detailed analytical treatment of the uncertainty function can be found in Refs. \cite{hu93a,anastopoulos95,hu95,koks97}.

\begin{figure}
\centering
    \scalebox{0.4}{\includegraphics{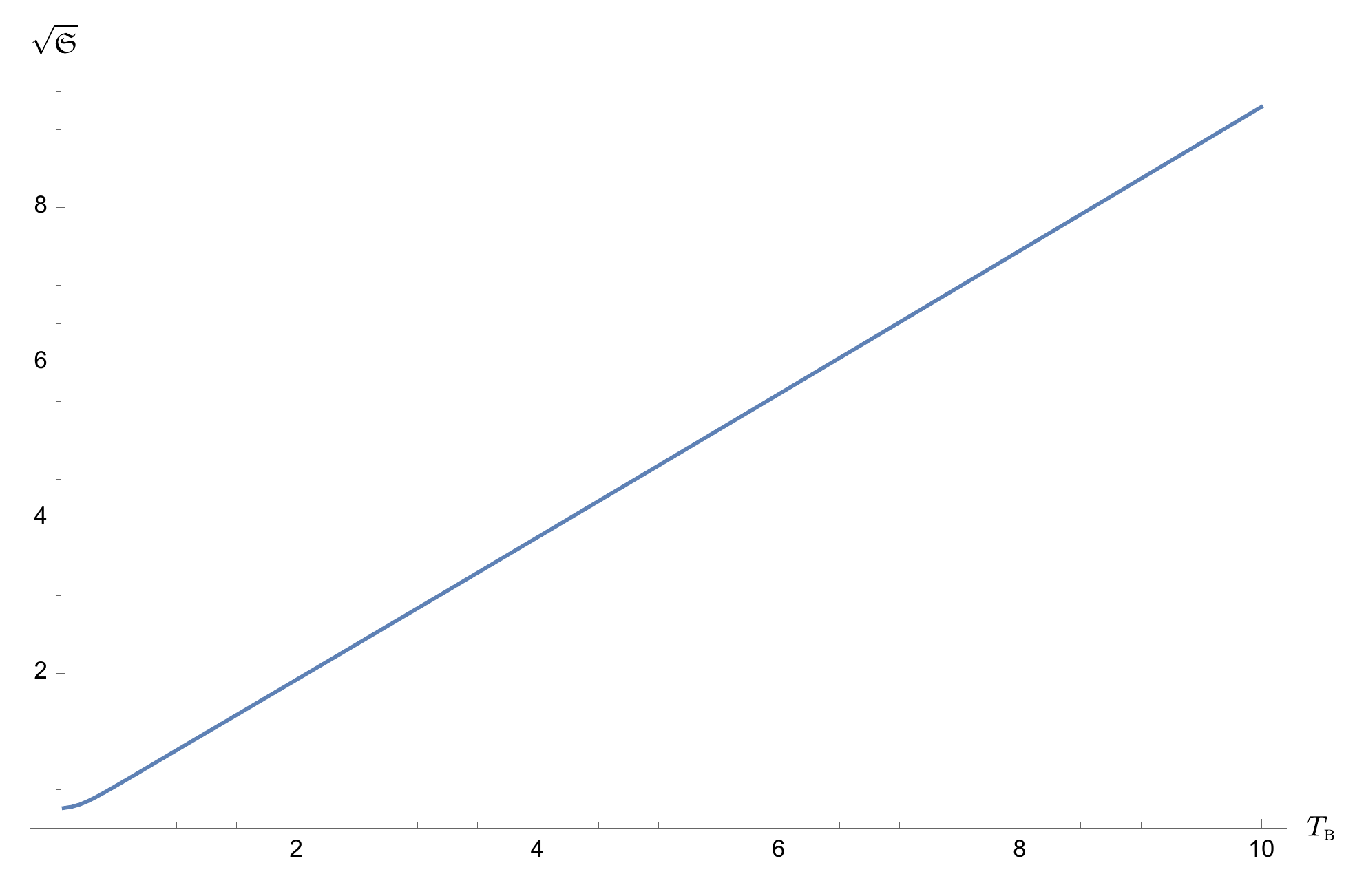}}
    \caption{The dependence of the uncertainty function $\sqrt{\mathfrak{S}(t)}$ at late times on the bath temperatures. We choose the time $t$ to be $t=10\omega_{\textsc{r}}^{-1}$ so that $\gamma t=3$, {i.e.} the regime where the relaxation is almost complete.
		We take the square root of the uncertainty function {to reveal} the linear trend at sufficiently high bath temperature where the equipartition theorem applies.
		We choose the same values for the other parameters used in Fig.~\ref{Fi:UnCent}.
		}
		\label{Fi:UnCentT}
\end{figure}

In Fig.~\ref{Fi:UnCentT}, we show the uncertainty function {as a function of the bath temperature }at late times.
It is clearly seen that at sufficiently high temperatures, $\sqrt{\mathfrak{S}}$ grows linearly with the bath temperature $T_{\textsc{b}}^{\vphantom{-1}}=\beta_{\textsc{b}}^{-1}$.
This can be understood {from} the equipartition theorem, {namely}
\begin{align}
	\frac{a}{2m}=\frac{\langle\hat{P}^{2}\rangle}{2m}&\sim\frac{T_{\textsc{b}}}{2}\,,&\frac{m\omega_{\textsc{r}}^{2}}{2}b=\frac{m\omega_{\textsc{r}}^{2}}{2}\langle\hat{Q}^{2}\rangle&\sim\frac{T_{\textsc{b}}}{2}\,,&c&\to0\,,&&\Rightarrow&\mathfrak{S}&\sim\omega_{\textsc{r}}^{2}T_{\textsc{b}}^{2}\,.
\end{align}
On the other hand, at very {low} temperatures, {where quantum effects become important, }we notice that the curve flattens.
This is a special feature we will see in the context of the effective temperature, discussed in the next section.

\section{Nonequilibrium quantum thermodynamics and uncertainty relations}\label{Sec:NEqThermodynamic}

The Robertson-Schr\"odinger uncertainty principle plays a distinguished role in nonequilibrium quantum thermodynamics of Gaussian open systems, because {the corresponding Gaussian state} can be uniquely defined by the second moments of the canonical variables, i.e. the building blocks of the uncertainty principle.
{In fact, in the following, we will see how the uncertainty principle shows naturally in the reduced system's density matrix and can be used to define an \textit{effective} temperature $\beta^{-1}_{\mathrm{eff}}$ of the system that includes the effects of the nonequilibrium evolution at strong coupling. Using these building blocks, we will continue to show how the uncertainty function is adopted by various definitions for the nonequilibrium thermodynamic functions. We will conclude this section by relating the uncertainty function to another nonequilibrium principle, the fluctuation-dissipation inequality, and comment on the transition to equilibrium and the emergence of a fluctuation-dissipation theorem.}

\subsection{Nonequilibrium partition function and effective temperature}\label{Sec:NEqPart}

Following~{Ref.~\cite{hsiang21}}, the density matrix operator of the reduced system at any time takes the form
\begin{align}\label{E:gfjfdefere}
	\hat{\rho}_{\textsc{s}}(\hat{Q},\hat{P},t)
	&=
	{2\sinh\frac{\vartheta(t)}{2}}\,
	\exp\biggl\{-{\vartheta(t)\,\tanh\frac{\vartheta(t)}{2}}\,\Bigl[a\,\hat{Q}^{2}+b\,\hat{P}^{2}-c\,\bigl\{\hat{Q},\,\hat{P}\bigr\}\Bigr]\biggr\}\notag\\
	&=\frac{1}{\mathcal{Z}_{\textsc{s}}}\,\exp\Bigl[-\beta_{\mathrm{eff}}(t)\hat{H}_{\mathrm{eff}}(t)\Bigr]\,,
\end{align}
where the nonequilibrium partition function $\mathcal{Z}_{\textsc{s}}(t)$ has been given by \eqref{E:kgkdtr},
\begin{align}
\label{E:rtvhd}
\mathcal{Z}_{\textsc{s}}(t)&=\frac{1}{2\sinh\frac{\vartheta(t)}{2}}\,,
&&\text{with}&\coth^{2}\frac{\vartheta(t)}{2}&=4\bigl[a(t)b(t)-c^{2}(t)\bigr].
\end{align}
We observe that the nonequilibrium partition function \eqref{E:rtvhd} has the same functional form as the one for the quantum harmonic oscillator in conventional thermodynamics. This implies that we may introduce an inverse effective temperature of the reduced system by
\begin{equation}\label{E:ujrtbr}
	\beta_{\mathrm{eff}}(t)=\frac{\vartheta(t)}{\omega_{\textsc{r}}}\,.
\end{equation}
In addition, we find that this partition function is related to the Robertson-Schr\"odinger uncertainty principle by
\begin{equation}
	\mathcal{Z}_{\textsc{s}}(t)=\sqrt{\mathfrak{S}(t)}\,,
\end{equation}
and thus
\begin{equation}
	\beta_{\mathrm{eff}}(t)=\frac{2}{\omega_{\textsc{r}}}\,\sinh^{-1}\frac{1}{2\sqrt{\mathfrak{S}(t)}}\,.\label{E:potrpj}
\end{equation}
This provides a bridge between the quantum uncertainty principle and nonequilibrium quantum thermodynamic relations.
It may not be too unexpected that the effective temperature $T^{\vphantom{-1}}_{\mathrm{eff}}(t)=\beta^{-1}_{\mathrm{eff}}(t)$ is independent of the squeeze parameter $\zeta$, even though {squeezing} accounts for parts of the dynamical features of the reduced system. In a sense, the uncertainty function, which is a manifestation of the uncertainty principle attributed to the noncommutativity in quantum physics, can be viewed as a measure of the overall quantum fluctuations, originated from the conjugated pair {of operators} minus their correlation.
From a phase space description in terms of the Wigner functions, we know that the squeezing may distort the quadratures that contribute to the uncertainty function, but it does not change its value.
The dependence of the nonequilibrium partition function, and the effective temperature {alike}, on the uncertainty function seems to pinpoint the central role of fluctuations in quantum thermodynamics, at least for Gaussian open systems as demonstrated here. Thus the effective temperature is more associated with the overall fluctuations, described by the uncertainty function, rather than the fluctuations encoded in only one of the canonical variables of the system, such as the momentum fluctuations in the form of the mean kinetic energy.
In mathematical terms, from the perspective of the nonequilibrium partition function $\mathcal{Z}_{\rm s}$, it is the product of the correlations that counts, not their weighted sum.

\paragraph{Effective temperature in a dynamical setting}
The effective temperature we introduced has more dynamical connotations than statistical ones. For the  initial state of the system we adopted, that is, the ground state of the harmonic oscillator, we easily find {that} the corresponding effective temperature vanishes, i.e. from taking the limit in Eq. \eqref{E:potrpj} we obtain $\displaystyle
\lim_{t\to0}\beta_{\mathrm{eff}}^{-1}(t)
	=
	{\lim_{x\to 0^+}\bigl[-\omega_{\textsc{r}}/\log(x)\bigr]=0}
$
since $\mathfrak{S}\to0$ for $a(0)=b(0)=1/2$.
As the interaction between the system and the bath proceeds, this effective temperature changes with time in a way similar to the uncertainty function because the former is a monotonic function of the latter.
Following the itemized discussions in the context of the covariance matrix elements {(see discussions at the end of Sec.~\ref{Sec:DynB})}, we note that, as the system approaches relaxation, the effective temperature gradually loses its dependence on the initial state, and it eventually inherits the statistical nature of the thermal bath distorted by {finite coupling strength}.
When equilibration is reached, {the} effective temperature will asymptotically saturate to a time-independent constant.
This constant in general is not equal to the bath temperature, except for the limiting case of vanishing system-bath coupling,
{where at late times
$
\mathfrak{S}\sim [\coth^2(\beta_{\textsc{b}}\omega_{\textsc{r}}/2)-1]/4
$
such that we have $T_{\mathrm{eff}}=T_{\textsc{b}}$.
}
{The similar situation is achieved at very high temperatures
$\beta_{\textsc{b}}\omega_{\textsc{r}}\ll 1$, where $\mathfrak{S}\sim(\beta_{\textsc{b}}\omega_{\textsc{r}})^{-2}-1/4$ is independent of the coupling strength (see also Sec.~\ref{Sec:FDIandRS}).
}
\begin{figure}
\centering
    \scalebox{0.39}{\includegraphics{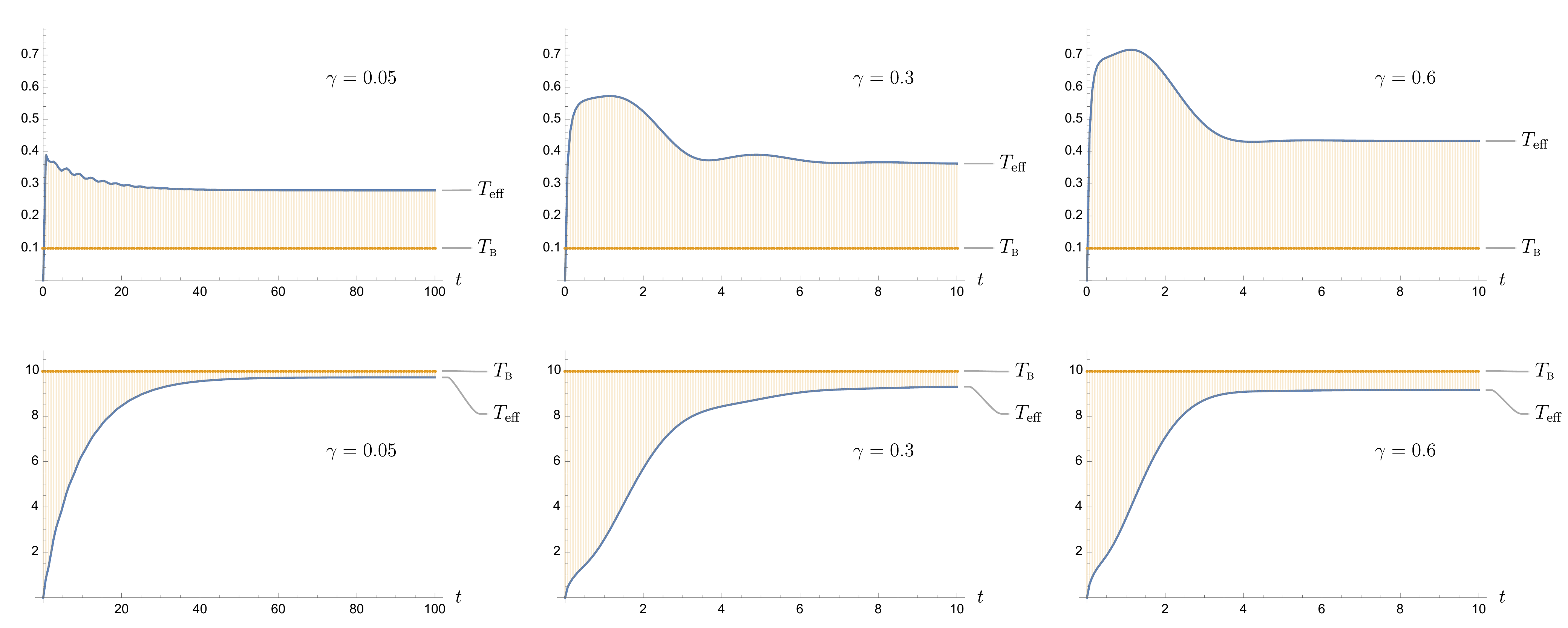}}
    \caption{We show the time dependence of the effective temperature, and the bath temperature is drawn as the reference. The top row corresponds to the low bath temperature cases, and the bottom row to the high bath temperature. Here the temperature is normalized to the oscillator frequency. The three columns represent $\gamma=0.05$, $\gamma=0.3$ and $\gamma=0.6$ from left to right. The cutoff parameter $\epsilon$ is $\epsilon=0.05$. We observe that the gap between the effective temperature $T_{\mathrm{eff}}$ and the bath temperature $T_{\textsc{b}}$ at late times decreases with weaker system-bath coupling. Thus they are best matched in the limits of high bath temperature and vanishing coupling strength, the common setting in the traditional thermodynamics.
		}\label{Fi:effectiveTemp}
\end{figure}

The temporal behavior of the effective temperatures for different damping constants {and bath temperatures} are shown in Fig.~\ref{Fi:effectiveTemp}.
{As} expected, the effective temperature shows a more dramatic difference from the bath temperature in the low bath temperature $T_{\textsc{b}}$ {limit (top row of Fig. \ref{Fi:effectiveTemp})}, {in combination with strong system-bath coupling (increasing from left to right in Fig.~\ref{Fi:effectiveTemp})}.
For comparison, the difference becomes marginal at high bath temperatures (bottom row of Fig.~\ref{Fi:effectiveTemp}), {and weak coupling (decreasing coupling from right to left in Fig.~\ref{Fi:effectiveTemp}).
Perhaps the most interesting regime is low bath temperatures and strong system+bath coupling (top right plot in Fig. \ref{Fi:effectiveTemp}). As has been pointed out in~\cite{hsiang21}, the fact that the curve of the effective temperature in the limit of zero bath temperature flattens out to a nonzero, positive value, is a consequence of finite coupling and evidence of nonvanishing quantum entanglement between the system and the bath.

In the high-bath-temperature regime, the late-time values of the effective temperature do not differ much from the value of the bath temperature, because the damping constant plays a subdominant role, compared to the bath temperature.
This leads to an intriguing phenomenon {connected to the regularization scheme}.
As we pointed out earlier, the effective temperature is a function of the covariance matrix elements.
Due to the system-field interaction, the element describing the momentum uncertainty is not well defined {and needs to be regularized}.
When we evaluate \eqref{E:fnghj}, we will end up with an integral over the frequency variable $\kappa$.
This integral is logarithmically divergent, so we insert a convergent factor of the form $e^{-\kappa\epsilon}$ with $\epsilon^{-1}$ playing the role of the regularization scale, i.e., the cutoff frequency.
Alternatively, among various regularization schemes, we may implement regularization by setting the upper limit of the $\kappa$ integral to $\epsilon^{-1}$.
Thus, the momentum uncertainty has a term proportional to the logarithm of the cutoff frequency. {Similarly, }different regularization schemes typically generate a proportionality constant of the same order of magnitude, but not the same numerically value.
In addition, from the theoretical consideration, the choice of the regularization scale may not be explicit.
The configuration at hand may not have a suitable candidate on the scale of our interest, or may have more than one possibilities.
Thus, it leaves room for ambiguity {in the choice of the regularization}, in this case, for the numerical value of the momentum uncertainty, and in turn the effective temperature.
This ambiguity sometimes does not pose an issue due to its logarithmic dependence on the cutoff scale, but in the high-bath-temperature regime, the value of the effective temperature at late times, though close to the bath temperature, can be greater or less than it, as shown in Fig.~\ref{Fi:cutoff} {or the lower row of Fig. \ref{Fi:effectiveTemp}}.
{In particular, we note that this occurs for a cutoff that is coming closer to $\omega_{\textsc{r}}$ (right panel in Fig. \ref{Fi:cutoff}). For an increasing frequency-cutoff (decreasing $\epsilon$), the effect disappears and one once again enters the regime $T_{\mathrm{eff}}>T_{\textsc{b}}$.}
{We further note} that the cutoff dependence of the system quantities, resulting from interaction with the field, does not render them unphysical.
The scale ultimately will be fixed by the energy scale or other scales in experiments.
Finally, {the presence} of the cutoff scale is physically necessary to enforce the weaker form {(including the quantum average)} of the equal-time commutation relation for the canonical variables of the reduced system.
As shown in Appendix~\ref{App:RSUncertaintyPrinciple}, the weaker form \eqref{E:rotugbdf} depends on the unitarity of the reduced density matrix $\hat{\rho}_{\textsc{s}}$, i.e., $\operatorname{Tr}_{\textsc{s}}\hat{\rho}_{\textsc{s}}=1$.
Since for a Gaussian state, the covariance matrix elements are the constituents of the density matrix elements, dropping the cutoff-dependent terms in the covariance matrix elements will violate the unitarity, and thus, following in the derivation in Appendix~\ref{App:RSUncertaintyPrinciple}, impairs the Robertson-Schr\"odinger uncertainty principle. {Still, since we model the thermal bath by a massless quantum scalar field, we may find $T_{\mathrm{eff}}<T_{\textsc{b}}$ or $T_{\mathrm{eff}}>T_{\textsc{b}}$ at the high $T_{\textsc{b}}$ regime. It depends on the choice of regularization scheme and the cutoff parameter. This phenomenon would serve as a reminder that when one would like to interpret the relation between the effective temperature and the bath temperature, one needs to heed  the physical range of the system parameters and the implementation of regularization. }

\begin{figure}
\centering
    \scalebox{0.4}{\includegraphics{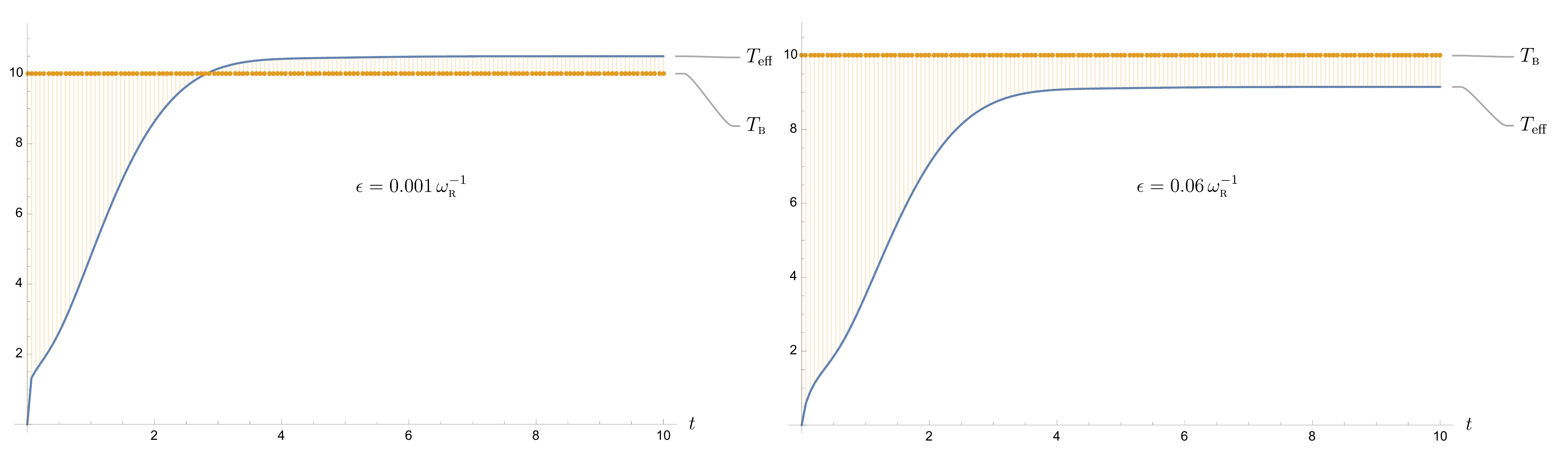}}
    \caption{We show the cutoff dependence of the effective temperature at late times, with the bath temperature as the reference. Its numerical value depends on the frequency cutoff and the regularization scheme. Here we choose $T_{\textsc{b}}=10\,\omega_{\textsc{r}}$, and pick the soft cutoff scheme, that is, inserting a convergent factor of the form $e^{-\kappa\epsilon}$ in the frequency integral over $\kappa$. The cutoff frequency is of the order $\epsilon^{-1}$. The effective temperature can be smaller than the bath's initial temperature if we choose a smaller cutoff frequency.}\label{Fi:cutoff}
\end{figure}
\subsection{Hamiltonian of mean force and Internal energy}

The density matrix \eqref{E:gfjfdefere} allows us to define~{\cite{hsiang21}} a nonequilibrium Hamiltonian of mean force $\hat{H}_{\mathrm{eff}}(t)$,
\begin{equation}
\label{Eq:EffectiveH}
	\hat{H}_{\mathrm{eff}}(t)=\omega_{\textsc{r}}\tanh\frac{\beta_{\mathrm{eff}}\omega_{\textsc{r}}}{2}\,\Bigl[a(t)\,\hat{Q}^{2}+b(t)\,\hat{P}^{2}-c(t)\,\bigl\{\hat{Q},\,\hat{P}\bigr\}\Bigr]\,.
\end{equation}
{In order to extract its physical meaning, it is interesting to compare Eq.~\eqref{Eq:EffectiveH} to other often used expressions. One obvious choice would be the system's Hamiltonian that is connected to its mechanical energy, i.e.}
\begin{equation}\label{E:gkfjbd}
	\hat{H}_{\textsc{s}}=\frac{m\omega^{2}_{\textsc{r}}}{2}\,\hat{Q}^{2}+\frac{1}{2m}\,\hat{P}^{2}\,.
\end{equation}
{The relation between $\hat{H}_{\mathrm{eff}}$ and $\hat{H}_{\textsc{s}}$ is established by means of a time-dependent squeezing transformation,}
\begin{equation}\label{E:hhnheersd}
	\hat{H}_{\mathrm{eff}}(t)=\hat{S}(\zeta(t))\,\hat{H}_{\textsc{s}}\,\hat{S}^{\dagger}(\zeta(t))\,.
\end{equation}
{Although} the {squeeze} operator $\hat{S}(\zeta)$ is unitary, it does not imply that the nonequilibrium Hamiltonian of mean force describes a unitary dynamics, because it does not correspond to the time evolution of the complete reduced dynamics.
{The} nonequilibrium Hamiltonian of mean force {in Eq. \eqref{E:hhnheersd}} also differs from the conventional Hamiltonian of mean force $\hat{H}_{\textsc{mf}}$, which is defined with respect to the bath temperature in the context of equilibrium quantum thermodynamics~{\cite{hsiang18,kirkwood35, lutz11, hanggi20}}.
The relationship between these two Hamiltonians of mean force is subtler and is not possible to establish until the reduced dynamics is fully relaxed:
The idea is that the late-time dynamics of the linear reduced system is independent of its initial state, so we will have the same density matrix operator in the final equilibrium state. This implies that the {operator} $\beta_{\mathrm{eff}}\hat{H}_{\mathrm{eff}}$ will be equal to $\beta_{\textsc{b}}\hat{H}_{\textsc{mf}}$ in this asymptotic regime.
{Recalling that the} covariance matrix element $c(t)$ will vanish in the limit $t\to\infty$, {we find that the} equilibrium Hamiltonian of mean force $\hat{H}_{\textsc{mf}}$ is given by
\begin{equation}
	\hat{H}_{\textsc{mf}}=\frac{\beta_{\mathrm{eff}}(\infty)}{\beta_{\textsc{b}}}\,\omega_{\textsc{r}}\tanh\frac{\beta_{\mathrm{eff}}(\infty)\omega_{\textsc{r}}}{2}\,\Bigl[a(\infty)\,\hat{Q}^{2}+b(\infty)\,\hat{P}^{2}\Bigr]\,.
\end{equation}
In contrast to the system's Hamiltonian {$\hat{H}_{\textsc{s}}$}, both Hamiltonians of mean force contain contributions due to the system-bath interaction~\cite{hsiang17}.
In the limit of vanishing system-bath coupling, it can be shown that both Hamiltonians of mean force will approach the system's Hamiltonian.
Later we will see it may be futile to identify the contributions from the system, the bath or interaction, {separately}.
{These uncommon features may be attributable to unseparable} entanglement between the system and the bath.

{By implication, the} equilibrium state of the system after relaxation is not necessarily a Gibbs state, so the system does not inherit the bath temperature, and does not {enjoy a universal temperature independence of the details of the system}.
{A more unsettling consequence of the above discussion indicates that we are now confronted with ambiguous definitions of the internal energy. }
For example, $\langle\hat{H}_{\textsc{s}}\rangle$ and
}
$\mathcal{U}_{\textsc{s}}=\frac{\partial}{\partial\beta_{\mathrm{eff}}}\bigl(\beta_{\mathrm{eff}}\mathcal{F}_{\textsc{s}}\bigr)
$,
{with the nonequilibrium free energy}
$\mathcal{F}_{\textsc{s}}(t)=-\beta_{\mathrm{eff}}^{-1}(t)\ln\mathcal{Z}_{\textsc{s}}(t)$,
give distinct results at strong coupling~\cite{hsiang17,hsiang18}.
Another example is the entropy $\mathcal{S}$. At strong coupling the {following} two definitions of the entropy,
$-\operatorname{Tr}\hat{\rho}_{\textsc{s}}\ln\hat{\rho}_{\textsc{s}}
$ and
$
\beta_{\textsc{b}}^{2}\,\partial\mathcal{F}_{\textsc{s}}/\partial\beta_{\textsc{b}}
$,
are not equivalent.

\paragraph{Internal energy}
{In order to assess these discrepancies, we will have a closer look at the systems's internal energy.}
{We first note that the internal energy $\mathcal{U}_{\textsc{s}}$ is connected to the effective Hamiltonian per definition via}
\begin{equation}\label{E:hgjdfgw}
	\mathcal{U}_{\textsc{s}}(t)=\langle\hat{H}_{\mathrm{eff}}\rangle=\frac{\omega_{\textsc{r}}}{2}\,\coth\frac{\beta_{\mathrm{eff}}\omega_{\textsc{r}}}{2}=\frac{\omega_{\textsc{r}}}{2}\,\sqrt{1+4\mathfrak{S}}=\omega_{\textsc{r}}\,\sqrt{ab-c^{2}}
	\,,
\end{equation}
{where we explicitly show the different parametrization of the expression of internal energy.}
{For comparison, we find that the system's ``mechanical energy'' can be written as}
\begin{align}
	\langle\hat{H}_{\textsc{s}}\rangle&=\frac{a}{2m}+\frac{m\omega_{\textsc{r}}^{2}}{2}\,b=\mathcal{U}_{\textsc{s}}\,\cosh2\eta\,
\end{align}
{and we recall that $\eta$ is the modulus of the squeeze parameter.
Importantly, we recognize that, while the internal energy $\mathcal{U}_{\textsc{s}}$ depends explicitly on the uncertainty function, i.e. a multiplication of the covariance matrix elements {($ab-c^2$)}, the average mechanical energy $\langle\hat{H}_{\textsc{s}}\rangle$ is given by a weighted sum of the covariance matrix elements [see discussion at the beginning of Sec.~\ref{Sec:NEqPart}].
}
{By} the inequality between the arithmetic and the geometric mean, we {further} find
\begin{align}
	\langle\hat{H}_{\textsc{s}}\rangle\geq\omega_{\textsc{r}}\sqrt{ab}\geq\omega_{\textsc{r}}\,\sqrt{ab-c^{2}}=\mathcal{U}_{\textsc{s}}\geq\frac{\omega_{\textsc{r}}}{2}\,.\label{E:uutgkh}
\end{align}
Hence $\mathcal{U}_{\textsc{s}}$ is bounded from above by $\langle\hat{H}_{\textsc{s}}\rangle$, and has a lower bound $\omega_{\textsc{r}}/2$, a consistent consequence of the vacuum fluctuations.
The first equality applies when
\begin{equation}
	\frac{a}{2m}=\frac{m\omega_{\textsc{r}}^{2}}{2}\,b,
\end{equation}
that is, when the virial theorem applies. In the context of open systems, it refers to vanishingly weak system-bath coupling.
The {last} equality is saturated in the limit of zero temperature and extremely weak coupling with the bath. This simple analysis shows that the equivalence between $\langle\hat{H}_{\textsc{s}}\rangle$ and $\mathcal{U}_{\textsc{s}}$ is established when the system-bath coupling is weak and the system has equilibrated. These are the typical settings in conventional thermodynamics.

In the high-temperature limit $\beta_{\mathrm{eff}}\omega_{\textsc{r}}\ll1$, we have
\begin{align}
	\mathcal{U}_{\textsc{s}}=\frac{\omega_{\textsc{r}}}{2}\,\coth\frac{\beta_{\mathrm{eff}}\omega_{\textsc{r}}}{2}&\geq\frac{1}{\beta_{\mathrm{eff}}}\,.\label{E:lejori}
\end{align}
This inequality results from the functional behavior of the $\coth$-function, and the equality is asymptotically satisfied when $\beta_{\mathrm{eff}}\to0$.

\begin{figure}
\centering
    \scalebox{0.36}{\includegraphics{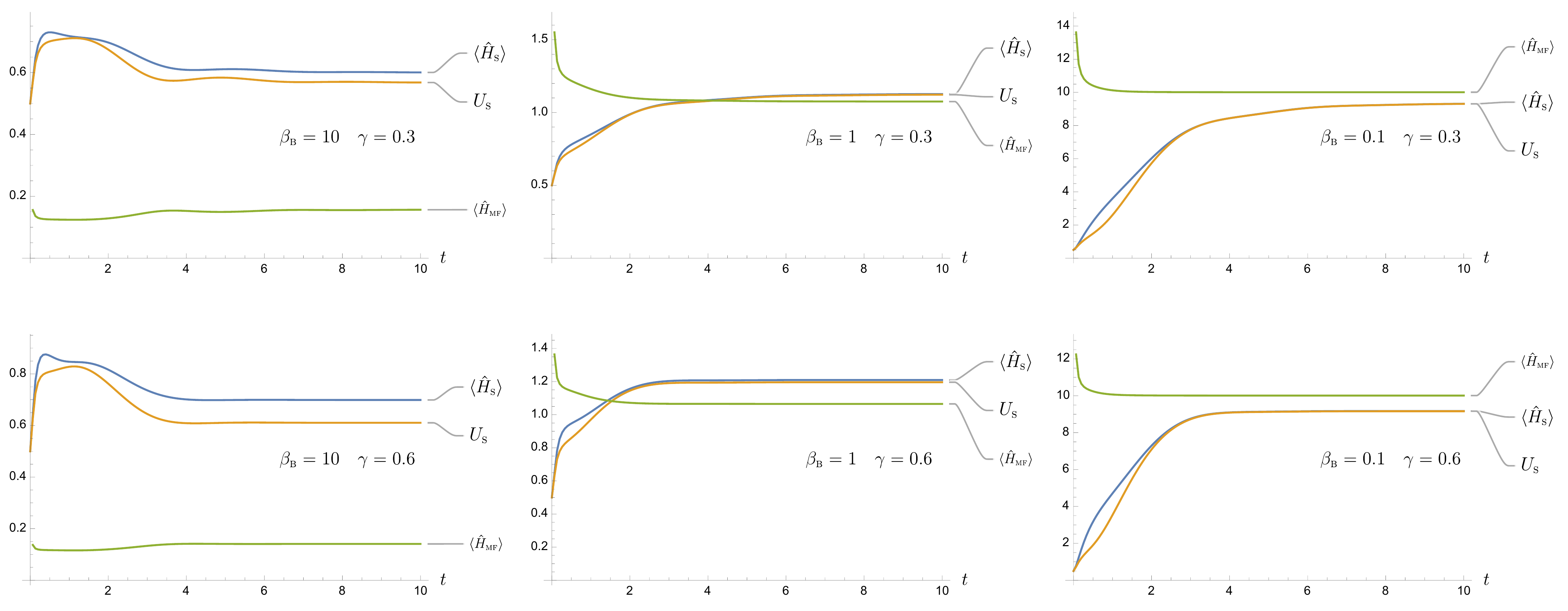}}
    \caption{The time dependence of various candidates of internal energy. The top row corresponds to the case $\gamma=0.3$, and the bottom row for $\gamma=0.6$. Here the temperature is normalized to the oscillator frequency. The three columns represent $\beta_{\textsc{b}}=10$, $\beta_{\textsc{b}}=1$ and $\beta_{\textsc{b}}=0.11$ from left to right, respectively corresponding to the high-, medium-, and low-temperature regime. Observe that $\langle\hat{H}_{\textsc{s}}\rangle$ is always greater or equal to $\mathcal{U}_{\textsc{s}}=\langle\hat{H}_{\mathrm{eff}}\rangle$ and the difference between them is smaller with higher bath temperature or weak system-bath coupling. By contrast, $\langle\hat{H}_{\textsc{mf}}\rangle$ is not grouped well together with the other two internal energies. This is related to the behavior of $\beta_{\mathrm{eff}}/\beta_{\textsc{s}}$ shown in Fig.~\ref{Fi:effectiveTemp}.}\label{Fi:intEng}
\end{figure}

\paragraph{Nonequilibrium thermodynamic inequalities}

The inequalities \eqref{E:uutgkh} and \eqref{E:lejori}, though identical {in appearance} to the familiar inequalities in traditional thermodynamics, are nonequilibrium by nature. They hold at all times, irrespective of the coupling strength between the system and the bath. In particular, $\langle\hat{H}_{\textsc{s}}\rangle\geq\langle\hat{H}_{\mathrm{eff}}\rangle=\mathcal{U}_{\textsc{s}}$ at every moment during the nonequilibrium evolution. The inequality between $\langle\hat{H}_{\textsc{s}}\rangle$
and $\langle\hat{H}_{\mathrm{eff}}\rangle$ is rather tight for the current setting because the correlation $c$ between the canonical pair is typically much smaller than the uncertainty of the canonical variables. This is clearly seen in Fig.~\ref{Fi:intEng}, where we show the time evolution of three candidates of the internal energy,
$\langle\hat{H}_{\textsc{s}}\rangle,~\langle\hat{H}_{\mathrm{eff}}\rangle$ and $\langle\hat{H}_{\textsc{mf}}\rangle$.
To avoid misinterpretations, a few comments are in place.
The traditional Hamiltonian of mean force $\hat{H}_{\textsc{mf}}$ is defined and discussed in the realm of equilibrium thermodynamics at strong coupling, so in principle its expectation value is not time-dependent.
The apparent time-dependence here comes from the density matrix operator used to evaluate the expectation value.
Thus, when making contact with quantities defined in the traditional setting, only the late-time values are needed. Secondly, the nonequilibrium Hamiltonian of mean force {$\hat{H}_{\mathrm{eff}}$} itself is a function of time, but this results purely from time evolution, not from the action of an external agent. Hence a decomposition like
\begin{equation}
	\frac{d}{dt}\mathcal{U}_{\textsc{s}}(t)=\operatorname{Tr}\Bigl\{\dot{\hat{\rho}}_{\textsc{s}}(t)\hat{H}_{\mathrm{eff}}(t)\Bigr\}+\operatorname{Tr}\Bigl\{\hat{\rho}_{\textsc{s}}(t)\dot{\hat{H}}_{\mathrm{eff}}(t)\Bigr\}\,\stackrel{?}{=}\,\dot{Q}+\dot{W}\,.
\end{equation}
can be misleading.  It may not be correct to identify the first term on the righthand side as the time rate of heat flow and the second term as the time rate of work done, and interpret this equation as the first law.
For comparison, using the mechanical energy, we can show that
\begin{equation}\label{E:gndkjsd}
	\frac{d}{dt}\langle\hat{H}_{\textsc{s}}(t)\rangle=\mathcal{P}_{\xi}(t)+\mathcal{P}_{\gamma}(t)
\end{equation}
irrespective of the coupling strength, at any moment during the evolution. The quantities $\mathcal{P}_{\xi}(t)$ and $\mathcal{P}_{\gamma}(t)$ are, respectively, the powers delivered by the quantum fluctuations from the quantum-field bath and the frictional force, reaction of the quantum radiation field due to the oscillator-field coupling. The righthand side accounts for the energy exchange between the system and the field bath during the nonequilibrium evolution. In particular, it can be shown that when the system reaches the equilibrium state, the righthand side ceases.

Moreover, from Fig.~\ref{Fi:intEng}, we observe that the behavior of $\langle\hat{H}_{\textsc{mf}}\rangle$ is quite detached from $\langle\hat{H}_{\textsc{s}}\rangle$ and $\langle\hat{H}_{\mathrm{eff}}\rangle$. The {discrepancy} is particularly severe at lower temperatures and stronger coupling.
This is related to the behavior of the ratio $\beta_{\mathrm{eff}}/\beta_{\textsc{s}}$ because the bath temperature $\beta_{\textsc{b}}^{-1}$ {does} not faithfully reflect the fluctuation phenomena in the system convened by the system-bath coupling{, especially before equilibration}.
On the other hand,  at higher bath temperatures, these three internal energies approach each other, so there is no difference in the weak-coupling, high-temperature thermodynamics. Incidentally, in the middle plot of the top row in Fig.~\ref{Fi:intEng}, the curve of $\langle\hat{H}_{\textsc{mf}}\rangle$ seems to merge closer with the other two than in the high temperature case to the right.
This is an artifact of the crossover between $\beta_{\textsc{b}}$ and $\beta_{\mathrm{eff}}$ at the medium bath temperature {as} can be {seen} from Fig.~\ref{Fi:effectiveTemp}.

\paragraph{Heat capacity}


From the internal energy we may introduce the corresponding heat capacity by~\cite{hsiang21}
\begin{equation}
	C_{\textsc{s}}=-\beta_{\mathrm{eff}}^{2}\frac{\partial\mathcal{U}_{\textsc{s}}}{\partial\beta_{\mathrm{eff}}}=\frac{\beta^{2}_{\mathrm{eff}}\omega^{2}_{\textsc{r}}}{4}\,\operatorname{csch}^{2}\frac{\beta_{\mathrm{eff}}\omega_{\textsc{r}}}{2}\geq0\,.
\end{equation}
It takes the same form as the one  in traditional thermodynamics, except that the bath temperature is replaced by the system's effective temperature and that this heat capacity is time-dependent.
{We aim to express the (dynamical) heat capacity in terms of} the dispersion of $\hat{H}_{\mathrm{eff}}$.
We first recall
{that
$
\hat{\rho}_{\textsc{s}}
	=
	\hat{S}(\zeta)\hat{\rho}_{\vartheta}\hat{S}^{\dagger}(\zeta)
$
such that we obtain} by \eqref{E:hhnheersd} that
\begin{equation}
	\langle\hat{H}_{\mathrm{eff}}^{2}\rangle=\operatorname{Tr}\Bigl\{\hat{\rho}_{\textsc{s}}\,\hat{H}_{\mathrm{eff}}^{2}\Bigr\}=\operatorname{Tr}\Bigl\{\hat{\rho}_{\vartheta}\hat{S}^{\dagger}(\zeta)\hat{H}_{\mathrm{eff}}^{2}\hat{S}(\zeta)\Bigr\}=\operatorname{Tr}\Bigl\{\hat{\rho}_{\vartheta}\hat{H}_{\textsc{s}}^{2}\Bigr\}
	=\langle\hat{H}^{2}_{\textsc{s}}\rangle_{\beta_{\mathrm{eff}}}\,,
\end{equation}
where we defined the average $\langle\cdots\rangle_{\beta_{\mathrm{eff}}}=\operatorname{Tr}\{\hat{\rho}_{\vartheta}\cdots\}$. Next,  since $\hat{\rho}_{\vartheta}$ is given by \eqref{E:rjugfg} with the help of \eqref{E:ujrtbr}, that is,
\begin{equation}
	\hat{\rho}_{\vartheta}=\frac{1}{{\mathcal{Z}_{\textsc{s}}}}\,e^{-\beta_{\mathrm{eff}}\hat{H}_{\textsc{s}}}
\end{equation}
and $\hat{H}_{\textsc{s}}$ is independent of $\beta_{\mathrm{eff}}$, we readily find that
\begin{align}
	\frac{\partial\mathcal{U}_{\textsc{s}}}{\partial\beta_{\mathrm{eff}}}=\frac{\partial}{\partial\beta_{\mathrm{eff}}}\operatorname{Tr}\Bigl\{\hat{\rho}_{\vartheta}\,\hat{H}_{\textsc{s}}\Bigr\}&=-\langle\hat{H}^{2}_{\textsc{s}}\rangle_{\beta}+\langle\hat{H}_{\textsc{s}}^{\vphantom{2}}\rangle_{\beta}^{2}\,,
\end{align}
where
\begin{align}
	\frac{\partial {\mathcal{Z}}_{\textsc{s}}}{\partial\beta_{\mathrm{eff}}}=\frac{\partial}{\partial\beta_{\mathrm{eff}}}\operatorname{Tr}\Bigl\{e^{-\beta_{\mathrm{eff}}\hat{H}_{\textsc{s}}}\Bigr\}=-\operatorname{Tr}\Bigl\{e^{-\beta_{\mathrm{eff}}\hat{H}_{\textsc{s}}}\hat{H}_{\textsc{s}}\Bigr\}=-{\mathcal{Z}}_{\textsc{s}}\,\mathcal{U}_{\textsc{s}}\,.
\end{align}
We thus arrive at
\begin{equation}
	C_{\textsc{s}}(t)=\beta_{\mathrm{eff}}^{2}\langle\Delta\hat{H}^{2}_{\textsc{s}}\rangle_{\beta}=\beta_{\mathrm{eff}}^{2}\langle\Delta\hat{H}_{\mathrm{eff}}^{2}\rangle\,,
\end{equation}
via the same familiar manipulations used in the traditional thermodynamics. Thus this nonequilibrium heat capacity is still proportional to the internal energy fluctuations, now defined by the nonequilibrium Hamiltonian of mean force \eqref{E:hgjdfgw}.
It thereby provides a physical link between $\hat{H}_{\textsc{s}}$ and $\hat{H}_{\mathrm{eff}}$, where, again, the fundamental core is the uncertainty function.

\section{Fluctuation-dissipation inequality and Robertson-Schr\"odinger uncertainty}\label{Sec:FDIandRS}

Until now, we have elaborated on the special role of the Robertson-Schr\"odinger uncertainty principle in Gaussian open systems from the perspective of quantum thermodynamic quantities.
{We will now continue to explore the role of the uncertainty function in the equilibration process by exploring its connection to the fluctuation-dissipation theorem.
The {connecting principle} will be given by the so-called fluctuation-dissipation inequality which is valid over the full course of the equilibrium evolution. Interestingly, we will see that at least one additional assumption {is needed} for providing the full link between the uncertainty function, over the fluctuation-dissipation inequality, all the way down to the fluctuation-dissipation theorem at late times. It will become necessary to specify the bath Hamiltonian and define its concrete statistical properties which are then inherited by the system dynamics.}

We begin with {a brief derivation} of the fluctuation-dissipation inequality in a setting slightly more general than used in~\cite{fleming13}.

Given an operator $\hat{O}$ which is not necessarily Hermitian, we can always form a non-negative operator $\hat{O}^{\dagger}\hat{O}$ such that
\begin{equation}
	\langle\hat{O}^{\dagger}\hat{O}\rangle\geq0\,,
\end{equation}
as long as the expectation value is well-defined.
{For instance,} this immediately implies~\cite{ford88,fleming13}
\begin{equation}\label{E:kigbdffh}
	\int_{t'}^{t}\!\mathrm{d}s\mathrm{d}s'\;f^{*}(s)\,\langle\hat{o}^{\dagger}(s)\hat{o}(s')\rangle\,f(s')\geq0\,,
\end{equation}
if we assign
\begin{equation}
	\hat{O}(t)=\int^t_{t'}\!\mathrm{d}s\;\hat{o}(s)\,f(s)\,
\end{equation}
for any well-behaved complex function $f(s)$. In fact, we also readily have
\begin{equation}\label{E:ldghs}
	\int_{t'}^{t}\!{\mathrm{d}s\mathrm{d}s'}\;f^{*}(s)\,\langle\bigl\{\hat{o}^{\dagger}(s),\,\hat{o}(s')\bigr\}\rangle\,f(s')\geq0\,.
\end{equation}
because $\langle\hat{O}^{\dagger}\hat{O}\rangle\geq0$ implies $\langle\hat{O}\hat{O}^{\dagger}\rangle\geq0$.
It turns out useful to decompose the product $\hat{o}^{\dagger}(s)\hat{o}(s')$ into the sum of the commutator and anti-commutator of $\hat{o}(s)$ and $\hat{o}(s')$, and to write \eqref{E:kigbdffh} into
\begin{equation}\label{E:kgrt}
	\frac{1}{2}\int_{t'}^{t}\!{\mathrm{d}s\mathrm{d}s'}\;f^{*}(s)\,\langle\bigl\{\hat{o}^{\dagger}(s),\,\hat{o}(s')\bigr\}\rangle\,f(s')+\frac{1}{2}\int_{t'}^{t}\!{\mathrm{d}s\mathrm{d}s'}\;f^{*}(s)\,\langle\bigl[\hat{o}^{\dagger}(s),\,\hat{o}(s')\bigr]\rangle\,f(s')\geq0\,.
\end{equation}
Now we introduce the Hadamard, the Pauli-Jordan and the retarded Green's functions of the operator $\hat{o}$, i.e.
\begin{align}
	G_{\textsc{h}}^{(o)}(s,s')&\equiv \frac{1}{2}\langle\bigl\{\hat{o}^{\dagger}(s),\,\hat{o}(s')\bigr\}\rangle\,,&G_{\textsc{p}}^{(o)}(s,s')&\equiv i\,\langle\bigl[\hat{o}^{\dagger}(s),\,\hat{o}(s')\bigr]\rangle\,,\label{E:rgvsjhfgd}\\
	G_{\textsc{r}}^{(o)}(s,s')&\equiv i\,\theta(s-s')\,\langle\bigl[\{\hat{o}^{\dagger}(s),\,\hat{o}(s')\bigr]\rangle\,,
\end{align}
and rewrite Eq.~\eqref{E:kgrt} as
\begin{align}
	\int^t_{t'}\!\mathrm{d}s\!\int^t_{t'}\!\mathrm{d}s'\;f^{*}(s)\,G_{\textsc{h}}^{(o)}(s,s')\,f(s')\geq\frac{i}{2}\int^t_{t'}\!\mathrm{d}s\!\int^t_{t'}\!\mathrm{d}s'\;f^{*}(s)\,G_{\textsc{p}}^{(o)}(s,s')\,f(s')\,.\label{E:etiugfvg}
\end{align}
This is a general case of the fluctuation-dissipation inequality put forward in Ref. \cite{fleming13}.

Next {we} observe that if $\hat{o}(s)$ is Hermitian and $f$ is a real function of $s$, then the integrand on the righthand side of \eqref{E:etiugfvg} is odd in exchange of $s$ and $s'$, so the corresponding double integrals vanish, and we end up with
\begin{equation}
	\int^t_{t'}\!\mathrm{d}s\!\int^t_{t'}\!\mathrm{d}s'\;f(s)\,G_{\textsc{h}}^{(o)}(s,s')\,f(s')\geq0\,.
\end{equation}
This is the special case of \eqref{E:ldghs}. Thus we would like to at least require $f$ to be a complex function, or $\hat{o}$ to be non-Hermitian, or both. Otherwise the inequality \eqref{E:etiugfvg} is quite general, regardless of whether the dynamics associated with $\hat{o}$ or $f$ has an equilibrium state or not.

In the limiting case $t\to+\infty$, $t'\to-\infty$, and supposing that the Green functions $G^{(o)}(s,s')=G^{(o)}(s-s')$ is stationary, {the previous result in \eqref{E:etiugfvg} simplifies to}
\begin{equation}
	\int_{-\infty}^{\infty}\!\frac{\mathrm{d}\kappa}{2\pi}\;\lvert\tilde{f}(\kappa)\rvert^{2}\,\tilde{G}_{\textsc{h}}^{(o)}(\kappa)\geq\frac{i}{2}\int_{-\infty}^{\infty}\!\frac{\mathrm{d}\kappa}{2\pi}\;\lvert\tilde{f}(\kappa)\rvert^{2}\,\tilde{G}_{\textsc{p}}^{(o)}(\kappa)\,,
\end{equation}
where the Fourier transform of a function $f(t)$ is defined by
\begin{equation}
	\tilde{f}(\kappa)=\int_{-\infty}^{\infty}\!\mathrm{d}t\;e^{i\kappa t}\,f(t)\,.
\end{equation}
Thus, symbolically, we have
\begin{equation}\label{E:gjkurt}
	\tilde{G}_{\textsc{h}}^{(o)}(\kappa)\geq\frac{i}{2}\,\tilde{G}_{\textsc{p}}^{(o)}(\kappa)=\operatorname{Im}\tilde{G}_{\textsc{r}}^{(o)}(\kappa)>0\,,
\end{equation}
for $\kappa>0$. {Here, the last equality follows from the Kramers-Kronig relations for the causal Green function (see Ref. \cite{jackson99} and appendix \ref{App:KramersKronig}).
To extract the physical meaning for our case, we set $\hat{o}=\hat{\phi}_h(t)$ and define
\begin{align}
	G_{\textsc{p}}^{(\phi)}(\tau)&=-\frac{\partial}{\partial\tau}\Gamma^{(\phi)}(\tau)\,,&\tilde{G}_{\textsc{p}}^{(\phi)}(\kappa)&=i\,\kappa\,\tilde{\Gamma}^{(\phi)}(\kappa)=-2i\,\operatorname{Im}\tilde{G}_{\textsc{r}}^{(\phi)}(\kappa)\,.\label{E:rjtgfd}
\end{align}
Applying it to the Langevin equation of the internal dynamics of the harmonic {system} coupled to the field [Eq. \eqref{E:gkfha}], we find
\begin{align} \label{E:eotiuhdbs}
	\ddot{\hat{Q}}(t)+\omega_{\textsc{r}}^{2}\hat{Q}(t)+\frac{e^{2}}{m}\,\Gamma^{(\phi)}(t)\,\hat{Q}(0)=\frac{e}{m}\,\hat{\phi}_{h}(t)-\frac{e^{2}}{m}\int_{0}^{t}\!\mathrm{d}s\;\Gamma^{(\phi)}(t-s)\,\dot{\hat{Q}}(s)\,.
\end{align}
Then Eq.~\eqref{E:gjkurt} takes the form
$
\tilde{G}_{\textsc{h}}^{(\phi)}(\kappa)
	\geq
	|\frac{\kappa}{2}
	\tilde{\Gamma}^{(\phi)}(\kappa)|,
$
which represents the fact that as soon as there is any damping in the environment, that will cause fluctuations exceeding the $\propto\tilde{\Gamma}$ in their spectral magnitude.

{An} exact relation solely based on the fluctuation-dissipation inequality even at late times is not at hand.
Instead, {after the system equilibrated, one finds a fluctuation-dissipation \textit{equality} \cite{kubo66}.
It is interesting to note that, despite the similar nomenclature, the FDI does not generally evolve to an FDR at late times.
Indeed, in our case, the FDR reads
$
\tilde{G}_{\textsc{H}}^{(\phi)}(\kappa)
	=\coth[\beta_{\textsc{b}}\kappa/2]
	(\kappa/2)\tilde{\Gamma}^{(\phi)}(\kappa)
$.
Only in the limit of zero bath temperature ($\beta_{\textsc{b}}\to\infty$), will the FDI saturate its equality at late times and coincide with the FDR.
In any other more general case, the fluctuation-dissipation inequality serves as a lower bound for the bath fluctuations.}
In that sense, the fluctuation-dissipation inequality can be understood as the absolute quantum limit of the fluctuations in the system.
Mathematically, this means that {substituting} hermitian operators {with} the bath degrees of freedom, Eq.~\eqref{E:kigbdffh}, merely sets the playground for the system to evolve, but the specific thermodynamic behavior remains a prerogative of the statistical details of the bath.
Naively, this might be surprising, as one would probably assign the role {of describing the minimal uncertainty} to the Robertson-Schr\"odinger inequality.
Hence, it is interesting to explore the relation between the two inequalities.

To this end, we resume our discussion in Eq. \eqref{E:etiugfvg}.
We specify $f(s)=\theta(t-s)\,d_{2}(t-s)$ and assume a stationary Green's function of $\hat{\phi}$.
Without loss of generality, we let $t'=0$, and then Eq.~\eqref{E:etiugfvg} gives
\begin{align}
	&&&\int^t_{-\infty}\!\mathrm{d}s\!\int^t_{-\infty}\!\mathrm{d}s'\;\theta(s)\,d_{2}(s)\,G_{\textsc{h}}^{(\phi)}(s-s')\,\theta(s')\,d_{2}(s')\notag\\
	&&&\qquad\qquad\qquad\qquad\geq\frac{i}{2}\int^t_{-\infty}\!\mathrm{d}s\!\int^t_{-\infty}\!\mathrm{d}s'\;\theta(s)\,d_{2}(s)\,G_{\textsc{p}}^{(\phi)}(s-s')\,\theta(s')\,d_{2}(s')\,,\notag\\
	&\Rightarrow&&\int_{-\infty}^{\infty}\!\frac{\mathrm{d}\kappa}{2\pi}\;\tilde{G}_{\textsc{h}}^{(\phi)}(\kappa)\,\left\lvert\mathfrak{b}_{\kappa}(t)\right\rvert^{2}
	\geq
	\frac{i}{2}\int_{-\infty}^{\infty}\!\frac{\mathrm{d}\kappa}{2\pi}\;\tilde{G}_{\textsc{p}}^{(\phi)}(\kappa)\,\left\lvert{\mathfrak{b}_{\kappa}(t)}\right\rvert^{2}\,.\label{E:fgjbetr}
\end{align}
{Here, we introduce the shorthand notations}
\begin{align}
	\mathfrak{b}_{\kappa}(t)
		&=
		\int_{-\infty}^{t}\!\mathrm{d}s\;\theta(s)d_{2}(s)\,e^{-i\kappa s}\,,
		&\mathfrak{a}_{\kappa}(t)
		&=
		\int_{-\infty}^{t}\!\mathrm{d}s\;\theta(s)\dot{d}_{2}(s)\,e^{-i\kappa s}\,,
\end{align}
{By choosing $f(s)$ accordingly, equation \eqref{E:fgjbetr} would similarly apply for $\mathfrak{b}_{\kappa}\to\mathfrak{a}_{\kappa}$.}
{This is the generalization of Eq. \eqref{E:gjkurt} to arbitrary times. We aim to relate \eqref{E:fgjbetr} with the uncertainty function. For simplicity, we ignore the initial dynamics of the system and focus on the situation}  when the dynamics {is about to enter} the relaxation regime ($t\geq\gamma^{-1}$).
{We can then conveniently ignore} the contributions from the initial conditions {which} gradually become exponentially small without introducing much error.
Then the covariance matrix elements will be approximately given by
\begin{subequations}
\label{E:tjhgnbf1}
\begin{align}
	\langle\hat{Q}^{2}(t)\rangle&{\sim}
	\frac{e^{2}}{m^{2}}\int_{-\infty}^{\infty}\!\frac{d\kappa}{2\pi}\;\tilde{G}_{\textsc{h}}^{(\phi)}(\kappa)\,\lvert\mathfrak{b}_{\kappa}(t)\rvert^{2}\,,\\
	\langle\hat{P}^{2}(t)\rangle&{\sim}
	 e^{2}\int_{-\infty}^{\infty}\!\frac{d\kappa}{2\pi}\;\tilde{G}_{\textsc{h}}^{(\phi)}(\kappa)\,\lvert\mathfrak{a}_{\kappa}(t)\rvert^{2},\\
 	\frac{1}{2}\langle\bigl\{\hat{Q}(t),\,\hat{P}(t)\bigr\}\rangle&{\sim}
	\frac{e^{2}}{m}\int_{-\infty}^{\infty}\!
	\frac{d\kappa}{2\pi}\;
	\tilde{G}_{\textsc{h}}^{(\phi)}(\kappa)\,\mathfrak{b}^{\vphantom{*}}_{\kappa}(t)\mathfrak{a}^{*}_{\kappa}(t)\,,\\
	\langle\bigl[\hat{Q}(t),\,\hat{P}(t)\bigr]\rangle&{\sim}
	-i\,\frac{e^{2}}{m}\int_{-\infty}^{\infty}\!\frac{d\kappa}{2\pi}\;
	\tilde{G}_{\textsc{p}}^{(\phi)}(\kappa)\,
	\mathfrak{b}^{\vphantom{*}}_{\kappa}(t)\mathfrak{a}^{*}_{\kappa}(t)\,.
	\label{E:tjhgnbf}
\end{align}
\end{subequations}
%
{We note that, applying} the Cauchy-Schwarz inequality \cite{dlmf} implies
\begin{equation}\label{E:fjgbdf}
	\langle\hat{Q}^{2}(t)\rangle\langle\hat{P}^{2}(t)\rangle\geq\biggl[\frac{1}{2}\langle\bigl\{\hat{Q}(t),\,\hat{P}(t)\bigr\}\rangle\biggr]^{2}\,,
\end{equation}
{explicitly showing that the covariance matrix is positive semi-definite.}
{More interestingly, successively applying to $\langle\hat{Q}^{2}(t)\rangle\langle\hat{P}^{2}(t)\rangle$ the fluctuation-dissipation inequality Eq. \eqref{E:fgjbetr} and the Cauchy-Schwarz inequality}, we {we can restore the Robertson-Schrödinger inequality, i.e.}
\begin{align}
	\langle\hat{Q}^{2}(t)\rangle\langle\hat{P}^{2}(t)\rangle&=\frac{e^{4}}{m^{2}}\int_{-\infty}^{\infty}\!\frac{\mathrm{d}\kappa}{2\pi}\;\tilde{G}_{\textsc{h}}^{(\phi)}(\kappa)\,\lvert\mathfrak{b}_{\kappa}(t)\rvert^{2}\int_{-\infty}^{\infty}\!\frac{\mathrm{d}\kappa'}{2\pi}\;\tilde{G}_{\textsc{h}}^{(\phi)}(\kappa')\,\lvert\mathfrak{a}_{\kappa'}(t)\rvert^{2}\notag\\
	&\stackrel{\textsc{fdi}}{\geq}
	\frac{1}{4}
	\frac{e^{4}}{m^{2}}\int_{-\infty}^{\infty}\!\frac{\mathrm{d}\kappa}{2\pi}\;\,\tilde{G}_{\textsc{p}}^{(\phi)}(\kappa)\,\lvert\mathfrak{b}_{\kappa}(t)\rvert^{2}\int_{-\infty}^{\infty}\!\frac{\mathrm{d}\kappa'}{2\pi}\;\,\tilde{G}_{\textsc{p}}^{(\phi)}(\kappa')\,\lvert\mathfrak{a}_{\kappa'}(t)\rvert^{2}\notag\\
	&\stackrel{\textsc{cs}}{\geq}\frac{1}{4}\,\left\lvert \,\frac{e^{2}}{m}\int_{-\infty}^{\infty}\!\frac{\mathrm{d}\kappa}{2\pi}\;\tilde{G}_{\textsc{p}}^{(\phi)}(\kappa)\,\mathfrak{b}^{\vphantom{*}}_{\kappa}(t)\mathfrak{a}^{*}_{\kappa}(t)\right\rvert^{2}\,.\label{E:otuhg}
	\\\nonumber
	&=
	\frac{1}{4}
	\left|\langle[\hat{Q}(t),\hat{P}(t)]\rangle\right|^2\,,
\end{align}
for $t\geq\gamma^{-1}$.
{Hence, we find that} the fluctuation-dissipation inequality provides a potentially sharper bound {to the uncertainty of the canonical variables than the Robertson-Schr\"odinger inequality} as long as the {last} inequality {in Eq.~\eqref{E:otuhg}} is not {saturated}.
%
{
In fact, the Robertson-Schr\"odinger inequality, for the canonical variables $\hat{Q}$ and $\hat{P}$, {is given by} the trivial form $\langle[\hat{Q}(t),\hat{P}(t)]\rangle=i$ over the full course of the nonequilibrium evolution, as one would expect from traditional quantum mechanical considerations (see appendix \ref{App:RSUncertaintyPrinciple}).
Since this might not be obvious from the evolved expression in Eq.~\eqref{E:tjhgnbf}, let us consider the case of late times explicitly.
}

We return to Eq. \eqref{E:tjhgnbf1} and examine the $t\to\infty$ limit. The solutions to Eq.~\eqref{E:eotiuhdbs} are
\begin{subequations}
\begin{align}
	\hat{Q}(t)&=d_{1}(t)\,\hat{Q}(0)+\frac{d_{2}(t)}{m}\,\hat{P}(0)+\frac{e^{2}}{m}\int^{t}_{0}\!\mathrm{d}s\;d_{2}(t-s)\,\hat{\phi}_{h}(s)\,,\\
	\hat{P}(t)&=m\,\dot{d}_{1}(t)\,\hat{Q}(0)+\dot{d}_{2}(t)\,\hat{P}(0)+e^{2}\int^{t}_{0}\!\mathrm{d}s'\;\frac{\partial}{\partial t}d_{2}(t-s')\,\hat{\phi}_{h}(s')\,.
\end{align}
\end{subequations}
{We hence obtain for the average of the commutator}
\begin{align}
	\langle\bigl[\hat{Q}(t),\,\hat{P}(t)\bigr]\rangle&=\Bigl[d_{1}(t)\dot{d}_{2}(t)-\dot{d}_{1}(t)d_{2}(t)\Bigr]\,\langle\bigl[\hat{Q}(0),\,\hat{P}(0)\bigr]\rangle\notag\\
	&\qquad
	-i\,\frac{e^{2}}{m}\int_{0}^{t}\!\mathrm{d}s\!\int_{0}^{t}\!\mathrm{d}s'\;d_{2}(t-s)\,\frac{\partial}{\partial t}d_{2}(t-s')\,G_{\textsc{p}}^{(\phi)}(s-s')\,.\label{E:rgjf}
\end{align}
Next we make a change of variables $y=t-s$, and then, {defining $\mathfrak{d}_2(y)=\theta(y)d_2(y)$,} we may extend the lower limit of the $y$-integral to $-\infty$. Eq.~\eqref{E:rgjf} {then} becomes
\begin{align}
	\langle\bigl[\hat{Q}(t),\,\hat{P}(t)\bigr]\rangle&=i\,e^{-2\gamma t}-i\,\frac{e^{2}}{m}\int_{-\infty}^{t}\!\mathrm{d}y\!\int_{-\infty}^{t}\!\mathrm{d}y'\;\mathfrak{d}_{2}(y)\,\dot{\mathfrak{d}}_{2}(y')\,G_{\textsc{p}}^{(\phi)}(y-y')\,.
\end{align}
The first term on the righthand side comes from the contributions of the initial conditions, and thus is negligible as $t\to\infty$. When $t$ goes to infinity, we arrive at
\begin{align}\label{E:eiotgsfds}
	\langle\bigl[\hat{Q}(\infty),\,\hat{P}(\infty)\bigr]\rangle&=-i\,\frac{e^{2}}{m}\int_{-\infty}^{\infty}\!\mathrm{d}y\!\int_{-\infty}^{\infty}\!\mathrm{d}y'\;\mathfrak{d}_{2}(y)\,\dot{\mathfrak{d}}_{2}(y')\,G_{\textsc{p}}^{(\phi)}(y-y')\notag\\
	&=-i\,\frac{e^{2}}{m}\int_{-\infty}^{\infty}\!\frac{\mathrm{d}\kappa}{2\pi}\;
	(-i\kappa)\,\tilde{G}_{\textsc{p}}^{(\phi)}(\kappa)\,\lvert\tilde{\mathfrak{d}}_{2}(\kappa)\rvert^{2}\,.
\end{align}
Now since
\begin{equation}\label{E:roetihg}
	\tilde{\mathfrak{d}}_{2}(\kappa)=\frac{1}{-\kappa^{2}+\omega_{\textsc{r}}^{2}-\frac{e^{2}}{m}\tilde{G}^{(\phi)}_{\textsc{r}}(\kappa)}\,,
\end{equation}
we can show that
\begin{equation}
	\tilde{\mathfrak{d}}_{2}^{\vphantom{*}}(\kappa)-\tilde{\mathfrak{d}}_{2}^{*}(\kappa)=i\,\frac{2e^{2}}{m}\,\operatorname{Im}\tilde{G}_{\textsc{r}}^{(o)}(\kappa)\,\lvert\tilde{\mathfrak{d}}_{2}(\kappa)\rvert^{2}\,.
\end{equation}
Thus with the help of \eqref{E:rjtgfd}, Eq.~\eqref{E:eiotgsfds} reduces to
\begin{align}
	\langle\bigl[\hat{Q}(\infty),\,\hat{P}(\infty)\bigr]\rangle=2i\int_{-\infty}^{\infty}\!\frac{\mathrm{d}\kappa}{2\pi}\;\kappa\,\operatorname{Im}\tilde{\mathfrak{d}}_{2}(\kappa)&=2i\operatorname{Im}\,\int_{-\infty}^{\infty}\!\frac{\mathrm{d}\kappa}{2\pi}\;
	\kappa\,\tilde{d}_{2}^{\theta}(\kappa)
	=2i\operatorname{Im}\Bigl\{i\,\theta(0)\,d_{2}(0)\Bigr\}\,.\label{E:riutbw}
\end{align}
Now observe that $d_{2}(0)=1$ and $\theta(0)=1/2$, and we finally find
\begin{equation}
	\langle\bigl[\hat{Q}(\infty),\,\hat{P}(\infty)\bigr]\rangle=i\,.
\end{equation}
{We would have obtained the same result by inserting Eq. \eqref{E:roetihg} and evaluating the first integral in Eq. \eqref{E:riutbw} directly. The given derivation, however, is more general as it relies on the functional properties of the Green tensor only and can hence be applied to any non-Markovian damping kernel.}

\section{Conclusion}\label{Sec:Conclusion}

Thermodynamic uncertainty principles (TUP) make up one of the few rare anchors in the largely uncharted waters of nonequilibrium systems. Our goal in this work, as stated in the beginning, is to trace the uncertainties of thermodynamic quantities in nonequilibrium systems all the way to their quantum origins, namely, to the (microscopic) quantum uncertainty principles (QUP).

The baseline of (quantum) thermodynamic inequalities is deeply rooted in the quantum-mechanical properties such as non-commutativity, interference and entanglement. Only on top of that come additional statistical effects of various sources. In this paper we have focused on the significance of non-commutativity in the form of the Robertson-Schr\"odinger uncertainty principle. This principle is of particular significance in a Gaussian system, because it can be expressed by the elements of the covariance matrix, which fully and uniquely determine the state of the thermodynamic open system. On the basis of previous work~\cite{hsiang21}, we highlight the role of the Robertson-Schr\"odinger uncertainty principle in the system's density matrix and the subsequently defined thermodynamic quantities such as the effective temperature.

%
During the course of the system's evolution, the {effective temperature} reflects the quantum-fluctuation dynamics of the system.
When the system-bath interaction is not weak, {the interaction between system and bath} will leave the imprint on the system dynamics. Noticeably, after the system comes to equilibrium, the effective temperature of the system is not equal to the bath temperature, and the disparity increases with stronger interaction strength. Furthermore, the equilibrium state of the system deviates from the Gibbs form. This necessarily implies that whatever effective temperature the system assumes, it is destined to be non-universal and leaves room for  specificity  in the precise physical meaning of its definition.

{Only} in the case of the Gaussian open systems, however, {where the trace of the density matrix is fixed by some form of a Gibbs state [see Eq. \eqref{E:rjugfg}],} would a judicious definition of the partition function {appear rather naturally}, {which} mitigates ambiguities in introducing the effective temperature.
Unique to {Gaussian systems interacting with a passive environment} is that the {resulting} thermodynamic quantities are solely dependent on the uncertainty function.
The approach based on the nonequilibrium partition function {then} leads to a notion of the system's internal energy and evokes an effective Hamiltonian $\hat{H}_{\mathrm{eff}}$ with respect to the effective temperature of the system $\beta^{-1}_{\mathrm{eff}}$, rather than with respect to the bath temperature $\beta^{-1}_{\textsc{b}}$.
This internal energy is shown to coincide with the expectation value of $\hat{H}_{\mathrm{eff}}$.

With this condition we then we compare the {result of the internal energy}  with two different, but equally plausible definitions of internal energies in the nonequilibrium setting at strong coupling.
On one hand, a customary example is the expectation value of the system Hamiltonian $\hat{H}_{\textsc{s}}$, which - in the classical mind - would correspond to the system's mechanical energy.
On the other hand, we have the expectation value of the Hamiltonian of mean force $\hat{H}_{\textsc{mf}}$ with respect to the bath temperature $\beta^{-1}_{\textsc{b}}$.
Interestingly, we observe that the {internal energy $\mathcal{U}_{\textsc{s}}=\langle\hat{H}_{\mathrm{eff}}\rangle$} is always bounded from above by the expectation value of the system Hamiltonian $\hat{H}_{\textsc{s}}$.
{Since the derivative with respect to time of} $\langle\hat{H}_{\textsc{s}}\rangle$ is the rate of energy exchanged between the system and the bath, the difference from the bound seems to be related to the energy cost when the system-bath entanglement is established due to the interaction.
We also find that in this nonequilibrium formulation, the heat capacity remains proportional to the uncertainty of the Hamiltonian $\hat{H}_{\mathrm{eff}}$, and the proportionality factor is given by the effective temperature of the system.

Lastly, returning to the equilibration process of the system, {we find that} the fluctuation-dissipation inequality, i.e. a nonequilibrium inequality on the magnitude of the respective fluctuation and dissipation kernels in the system, {can lead to a} fluctuation-dissipation \textit{equality} {in the zero-temperature limit} at late times.
{However, it} turns out that, in general, the fluctuation-dissipation inequality cannot by itself resolve into the fluctuation-dissipation equality at late times, but this is the case only for the absolute zero-temperature quantum bound.
Only with additional information on the bath spectral density can one  deduce the famous spectral $\coth(\beta\omega)/2$ behavior.
More interestingly, the fluctuation-dissipation inequality reproduces the Robertson-Schr\"odinger inequality at all times and can even serve as an upper, more precise bound.
Especially the RSI could provide viable information for fluctuation-limited experimental investigations.\\


\noindent {\bf Acknowledgements}
D.R. thanks Francesco Intravaia, Kurt Busch, and Markus Krutzik for enlightening discussions and gratefully acknowledges support from the German Space Agency (DLR) with funds provided by the Federal Ministry for Economic Affairs and Climate Action (BMWK) under grant number 50MW2251. J.-T. Hsiang is supported by the Ministry of Science and Technology of Taiwan, R.O.C. under Grant No.~MOST 110-2811-M-008-522.

\appendix

\section{Kramers-Kronig relation for the Green's function}
\label{App:KramersKronig}

From the definition of $G_{\textsc{h}}^{(o)}(s,s')$ and $G_{\textsc{p}}^{(o)}(s,s')$ of a non-Hermitian operator $\hat{o}$ in \eqref{E:rgvsjhfgd}, we obtain
\begin{align}
	G_{\textsc{h}}^{(o)\dagger}(s,s')&=G_{\textsc{h}}^{(o)}(s',s)=G_{\textsc{h}}^{(o)}(s,s')\,,&G_{\textsc{p}}^{(o)\dagger}(s,s')&=-G_{\textsc{p}}^{(o)}(s',s)\,,
\end{align}
which in turn implies that if the Green's functions are stationary, then
\begin{align}
	\tilde{G}_{\textsc{h}}^{(o)\dagger}(\kappa)&=\tilde{G}_{\textsc{h}}^{(o)}(\kappa)\,,&\tilde{G}_{\textsc{p}}^{(o)\dagger}(\kappa)&=-\tilde{G}_{\textsc{p}}^{(o)}(\kappa)\,.
\end{align}
Thus $\tilde{G}_{\textsc{h}}^{(o)}(\kappa)$ is real, but $\tilde{G}_{\textsc{p}}^{(o)}(\kappa)$ is imaginary. Next consider $\tilde{G}_{\textsc{r}}^{(o)}(\kappa)$, and we have
\begin{align}
	\tilde{G}_{\textsc{r}}^{(o)}(\kappa)
	&=
	\int_{-\infty}^{\infty}\!d\tau\;e^{i\kappa\tau}\,\theta(\tau)\,G_{\textsc{p}}^{(o)}(\tau)
	\\\nonumber
	&=
	i\int_{-\infty}^{\infty}\!\frac{d\omega}{2\pi}\;\frac{\tilde{G}_{\textsc{p}}^{(o)}(\omega)}{\omega-\kappa+i\,\epsilon}
	\\\nonumber
	&=
	\int_{-\infty}^{\infty}\!\frac{d\omega}{2\pi}\;\tilde{G}_{\textsc{p}}^{(o)}(\omega)\Bigl[i\,\mathfrak{P}\Bigl(\frac{1}{\omega-\kappa}\Bigr)-\pi\,\delta(\kappa-\omega)\Bigr]\,.
\end{align}
where we have added a convergent factor to make the integration well defined, and $\mathfrak{P}(\cdots)$ represents the principal value. This implies
\begin{equation}
	\operatorname{Im}\tilde{G}_{\textsc{r}}^{(o)}(\kappa)=\frac{i}{2}\,\tilde{G}_{\textsc{p}}^{(o)}(\kappa)\,.
\end{equation}

\section{The Robertson-Schr\"odinger uncertainty principle}\label{App:RSUncertaintyPrinciple}
The weaker commutation relation $\langle[\hat{Q}(t),\hat{P}(t)]\rangle=i$ still hold at an arbitrary $t$ for the open systems. Let $\hat{\rho}_{\textsc{s}}(t)$ be the reduced density matrix of the system. Then
\begin{align}\label{E:rotugbdf}
	\langle[\hat{Q}(t),\hat{P}(t)]\rangle&=\operatorname{Tr}_{\textsc{s}}\Bigl\{\bigl[\hat{Q},\hat{P}\bigr]\,\hat{\rho}_{\textsc{s}}(t)\Bigr\}\notag\\
	&=\int\!dQdQ'\;\delta(Q-Q')\Bigl\{Q\Bigl[-i\,\frac{\partial}{\partial Q}\rho_{\textsc{s}}(Q,Q',t)\Bigr]\Bigr\}-\Bigl\{-i\,\frac{\partial}{\partial Q}\Bigl[Q\,\rho_{\textsc{s}}(Q,Q',t)\Bigr]\Bigr\}\notag\\
	&=i\int\!dQdQ'\;\delta(Q-Q')\rho_{\textsc{s}}(Q,Q',t)\notag\\
	&=i\,.
\end{align}
Thus from the Schwarz inequality we find
\begin{align}
	\langle\hat{Q}^{2}(t)\rangle\langle\hat{P}^{2}(t)\rangle&=\operatorname{Tr}_{\textsc{s}}\Bigl\{\hat{Q}^{2}\hat{\rho}_{\textsc{s}}(t)\Bigr\}\times\operatorname{Tr}_{\textsc{s}}\Bigl\{\hat{P}^{2}\hat{\rho}_{\textsc{s}}(t)\Bigr\}\notag\\
	&\geq\lvert\operatorname{Tr}_{\textsc{s}}\Bigl\{\hat{Q}\hat{P}\,\hat{\rho}_{\textsc{s}}(t)\Bigr\}\rvert^{2}\notag\\
	&=\lvert\operatorname{Tr}_{\textsc{s}}\Bigl\{\frac{1}{2}\bigl\{\hat{Q},\hat{P}\bigr\}\,\hat{\rho}_{\textsc{s}}(t)\Bigr\}+\operatorname{Tr}_{\textsc{s}}\Bigl\{\frac{1}{2}\bigl[\hat{Q},\hat{P}\bigr]\,\hat{\rho}_{\textsc{s}}(t)\Bigr\}\rvert^{2}\,.
\end{align}
Since the first term is real but the second term is imaginary, we note from
\begin{align}
	z&=x+i\,y\,,&&\text{with}&x&\in\mathbb{R}\,,&y&\in\mathbb{R}\,,&&\Rightarrow&\lvert z\rvert^{2}&=x^{2}+y^{2}\,,
\end{align}
and then we conclude the RS uncertainty relation
\begin{align}
	\langle\hat{Q}^{2}(t)\rangle\langle\hat{P}^{2}(t)\rangle\geq\frac{1}{4}\langle\bigl\{\hat{Q}(t),\hat{P}(t)\bigr\}\rangle^{2}+\frac{1}{4}\,.
\end{align}



\newpage

\bibliography{./extracted}

\bibliographystyle{./prstytitlenew}

\end{document}